\documentclass[12pt,twoside]{article}
\usepackage{ecit02-1}
\usepackage{amssymb}
\usepackage{latexsym}
\usepackage{graphicx}
\newcommand{\eqref}[1]{~(\ref{#1})}

\author[D. Fournier-Prunaret , R. Lopez-Ruiz ]{ Dani\`{e}le Fournier-Prunaret, Ricardo Lopez-Ruiz}
 
\address{\rm D. Fournier-Prunaret\\ \it
LESIA - DGEI\\
INSA\\
135 avenue de Rangueil\\
31077 Toulouse, France\\
{\rm e-mail:}  daniele.fournier@insa-tlse.fr\\[.2in]
\rm R. Lopez-Ruiz\\ \it DIIS--Facultad de Ciencias\\ Instituto de Biocomputacion y Fisica de Sistemas Complejos\\
Universidad de Zaragoza\\ 50009 Zaragoza,
Spain\\
{\rm e-mail:} rilopez@unizar.es }

\title[Basin bifurcations in a 2-D map]
{Basin bifurcations in a two-dimensional logistic map}

\date{}

\begin{document}

\maketitle \makeaddress \msc{Primary 39B12, Secondary 26A18,
39B52} 

\begin{abstract}
Previous works have been devoted to the study of two-dimensional noninvertible maps, obtained using a coupling between one-dimensional logistic maps. This paper is devoted to the study of a specific one, in order to complete previous results \cite{Lo1} \cite{Lo3}, regarding the evolution of basins and attractors, when considering the tool of critical manifolds.
\end{abstract}

\section{Introduction}

 Many biological and physical phenomena are due to a complex connectivity among complex elements; feedback mechanism seems to be a suitable way to understand such phenomena (see for instance \cite{Ka1}\cite{Ka2}\cite{Ka3}). The introduction of feedback through a control parameter leads to models involving global multiplicative coupling among maps. One of the most simple models can be given by using logistic maps. Within this context, two-dimensional noninvertible maps with a global multiplicative coupling between one-dimensional logistic maps are of great interest. The understanding of their quite complex behaviour can be a first step in the understanding of dynamics in higher dimensional systems. 
Previous works (see \cite{Ga}\cite{Lo1}\cite{Lo2}\cite{Lo3}) have already been devoted to such kind of maps. In this paper, our aim is to complete the understanding of dynamics of model (c) given in \cite{Lo1}.
 
The considered model is the noninvertible two-dimensional map $T$ defined by:
\begin{eqnarray}\label{1}
\left\{\begin{array}{cc}
x_{k+1}=\lambda (3{y_k} + 1) x_k (1-x_k)
\\ 
y_{k+1}=\lambda (3{x_{k+1}} + 1) y_k (1-y_k)
\end{array}\right. 	
\end{eqnarray}    

where $\lambda$ is a real control parameter, $x$ and $y$ are real state variables. The map \eqref{1} includes a time asymmetric feedback. Numerical studies show that the system \eqref{1} is stable when $\lambda\ \in[-1.5,1.19]$.

Section 2 is devoted to recalls about fixed points of the map\eqref{1} and their bifurcations. Section 3 concerns critical manifolds. Section 4 analyzes evolution of basins by considering crossing of basin boundary through critical curves. In section 5 we explain evolution of invariant closed curve issued from a fixed point and giving rise to $weakly\ chaotic\ ring\ (WCR)$, then to chaotic attractor.

\section{Fixed points}

As explained in \cite{Lo1}, $T$ possesses at most five different fixed points :
\begin{eqnarray}\label{2}
P_0 : x_0=0,\ y_0=0 	
\end{eqnarray}
\begin{eqnarray}\label{3}
P_1 : x_1=\frac{\lambda-1}{\lambda},\ y_1=0,\quad P_2 : x_2=0,\ y_2=\frac{\lambda-1}{\lambda} 	
\end{eqnarray}
\begin{eqnarray}\label{5}
P_3 : x_3=\frac{1-\sqrt{a}}{3},\ y_3=\frac{1-\sqrt{a}}{3},   
\textrm{where  }a=\frac{4\lambda-3}{\lambda},\ \lambda>\frac{3}{4} 	
\end{eqnarray}
\begin{eqnarray}\label{6}
P_4 : x_4=\frac{1+\sqrt{a}}{3},\ y_4=\frac{1+\sqrt{a}}{3},\ \lambda>\frac{3}{4}   
\end{eqnarray}

For $-1<\lambda<1$, $P_0$ is an attractive node and $P_1$ and $P_2$ are saddle points on the negative side of the axes, x--axis being $P_1$ unstable manifold and y--axis being $P_2$ unstable one. When $\lambda=1$, $P_0=P_1=P_2$. For $\lambda>1$, $P_0$ is a repulsive node and $P_1$ and $P_2$ are saddle points, but located on the positive side of the axes, which are now $P_1$ and $P_2$ stable manifolds (Figures \ref{fig6}--\ref{fig7}). When $\lambda =-1$, $P_0$ undergoes a flip bifurcation and a period doubling cascade occurs when $\lambda$ decreases. $P_3$ and $P_4$ exist when $\lambda>\frac{3}{4}$. When $\lambda=\frac{3}{4}$, $P_3=P_4$; for $\frac{3}{4}<\lambda<1$, $P_3$ is a saddle point and $P_4$ an attractive node, the whole diagonal segment between $P_3$ and $P_4$ is locus of points belonging to heteroclinic trajectories connecting the two fixed points. When $\lambda=1$, $P_4$ undergoes a Ne\"\i mark--Hopf bifurcation and becomes repulsive by giving rise to an attractive invariant closed curve ($ICC$) ($C$). Section 5 is devoted to the study of ($C$) evolution.

\section{Critical lines}

An important tool used to study non-invertible maps is that of critical manifold, which has been introduced by Mira in 1964 (see \cite{Mi1}\cite{Mi2} for more details). A non-invertible map is characterized by the fact that a point in the state space can possess different number of rank-1 preimages, depending where it is located in the state space. A critical curve LC, in the two-dimensional case, is the geometrical locus of points X in the state plane having two coincident preimages, $T^{-1}$(X), located on a curve $LC_{-1}$. It is recalled that the set of points $T^{-n}$(X) constitutes the rank--n preimages of a given point X. The curve LC verifies $T^{-1}$(LC)$\supseteq$$LC_{-1}$ or T($LC_{-1}$)=LC and separates the phase plane in two regions where each point has a different number of preimages (k in one region and k+2 in the other one, k $\in\mathbb{N}$). For the map \eqref{1}, $LC_{-1}$ is given by the cancellation of the Jacobian of the map.

For the map \eqref{1}, the critical manifold LC$_{-1}$ is given by four arcs (Figure \ref{fig1}):
\begin{eqnarray}\label{7}
LC_{-1a} : x=\frac{1}{2},\quad LC_{-1b} : y=\frac{1}{2},\quad LC_{-1d} : y=-\frac{1}{3}
\end{eqnarray}
\begin{eqnarray}\label{8}
LC_{-1c} : y=-\frac{1}{3}-\frac{1}{9\lambda x(1-x)}
\end{eqnarray}

and the critical manifold LC by (Figure \ref{fig2}):
\begin{eqnarray}\label{9}
LC_a : y=\frac{4\lambda}{9}(3x+1)(\frac{4x}{\lambda}-1)(1-\frac{x}{\lambda})
\end{eqnarray}
\begin{eqnarray}\label{10}
LC_b : y=\frac{\lambda}{4}(3x+1),x\leq\frac{5\lambda}{8}
\end{eqnarray}
\begin{eqnarray}\label{11}
LC_c=\left\{(x=-\frac{1}{3},y=0)\right\}\subset LC_a\cap LC_b	
\end{eqnarray}
\begin{eqnarray}\label{12}
LC_d=\left\{(x=0,y=-\frac{4\lambda}{9})\right\}\subset LC_a	
\end{eqnarray}
Let us call  $Q_c$=$(x=\frac{5\lambda}{8},y=\frac{\lambda}{4}(\frac{15\lambda}{8}+1))$, $Q_c$ is a cusp point of $LC_{b}$, indeed $LC_b$ is constituted of two merged critical half straight lines ($x\ \in\ ]-\infty,\ \frac{5\lambda}{8}]$), which explains the fact that, by crossing $LC_b$, one goes from $Z_4$ area to $Z_0$ one, or from $Z_2$ to $Z_2$ (cf. Figure \ref{fig2}). Let us remark that $LC_{a}\cap LC_{b}=\left\{Q_c,\ (x=-\frac{1}{3},y=0)\right\}$.

The LC curves separate the state space in different areas $Z_i,\ i=0,2,4$ (Figure \ref{fig2}), where each point possesses $i$ distinct rank--1 preimages. Figure \ref{fig2a}  shows the state space as a foliated one, each sheet corresponding to the existence of a rank--1 preimage. This foliation looks like those given for parameter plane, for instance in \cite{Mi4}. Previous representations of such foliation related to the existence of LC curves can be found in \cite{Mi5}.

\section{Evolution of basins}

The basin of an attractor $B$ is defined as the set of all initial conditions that converge towards an attractor at finite distance when the number of iterations of $T$ tends towards infinity. The set $B$ verifies : $T^{-1}(B)=B$ and $T(B)\subseteq B$. We call $immediate\ basin$ the largest connected part of a basin containing the attracting set.
In this section, we study the evolution of basin of attractors when parameter $\lambda$ is modified. The situations are shown in Figures \ref{fig3} to \ref{fig17}. Using the terminology and the results of \cite{Mi1}\cite{Mi2}, our aim is to explain how basins become fractal, nonconnected or multiply connected in the case of \eqref{1}. 

\begin{itemize}
	\item Figure \ref{fig4} shows the first modification of the basin with the creation of a $bay$ H, due to the crossing of $LC_a$ through the basin boundary. The $bay$ H is a rank--1 preimage of the area $H_{-1}$, which is located inside $Z_2$, when $\lambda$ increases from the value 1.  
\end{itemize}
 
\begin{itemize}
	\item The second modification consists in a bifurcation $connected\ basin\ \leftrightarrow\ nonconnected$ $basin$ when $\lambda$ decreases from the value 1. Islands are created (cf. Figures \ref{fig6}-\ref{fig7}) after the bifurcation concerning the fixed points $P_0,\ P_1,\ P_2$. Indeed, when $\lambda$ decreases from the value 1, $P_1$ and $P_2$, located on positive part of axes, go to negative part of axes; the stable manifolds of these two saddle points is a part of the basin boundary; when the two points are on the negative part of axes, a tongue is created in the area ($x<0$, $y<0$) of the state space and crosses through $LC_a$ and $LC_b$ curves to penetrate inside $Z_2$ area, which is located above $LC_b$; it gives rise to a sequence of infinitely many preimages of any rank of this part of the tongue: these are the islands, the basin becomes non connected.
\end{itemize}

\begin{itemize}
	\item The third modification of the basin is due to a sequence of $aggregations\ of\ islands$ (Figures \ref{fig9}-\ref{fig11}). The first aggregation occurs when $LC_a$ and $LC_b$ curves become tangent in points $Q_a$ and $Q_b$ at the basin boundary (Figure \ref{fig9a}), rank--1 preimages of $Q_a$ and $Q_b$, $Q_a^{-1}$ and $Q_b^{-1}$, correspond to the aggregation of two islands with the immediate basin. The other preimages of $Q_a$ and $Q_b$ give rise to other aggregations. This first aggregation is followed by infinitely many new aggregations due to tangencies between critical curves and the immediate basin boundary when $\lambda$ decreases; such phenomena are explained more precisely in \cite{Mi3}. The important point to note is that a crossing between critical curves and basin boundary makes some domains change $Z_i$ area, so it makes appear or disappear preimages of the considered domains. The sequence of aggregations is nearly finished in Figure \ref{fig11}. The basin size increases when the parameter $\lambda$ decreases and tongues corresponding to island ends disappear when crossing critical curves of different rank (Figure \ref{fig12}). In Figure \ref{fig13}, the basin boundary is smooth. 
	When $\lambda$ decreases, being negative, an inverse process gives rise to  new appearance of tongues (Figure \ref{fig14}). 
\end{itemize}

\begin{itemize}
	\item Another modification corresponds to a bifurcation $connected\ basin$ $\leftrightarrow$ $multiply-connected\ basin$, that means basin with holes inside. The appearance of these holes is also related to crossing of basin boundary through critical curves; for instance, the hole in the middle of the basin in Figure \ref{fig15} is due to the crossing of $LC_a$ near its left extremum; it makes appear an area in $Z_2$ region and the preimages of any rank of this area gives rise to infinitely many holes inside the basin; these holes accumulate along the basin boundary. When $\lambda$ decreases, the size of holes increases (Figure \ref{fig16}). Some holes can open and be changed in $bays$ (cf. \cite{Mi1}\cite{Mi2}), this is the case in Figure \ref{fig17}, the basin boundary becomes fractal.
\end{itemize}

\section{Evolution of attractors}

This section is devoted to the study of evolution of attractors when $\lambda$ increases. This evolution can involve relations with the attractor basin or homoclinic and heteroclinic bifurcations.

\begin{itemize}
	\item When $\lambda$= 1, the attractive focus undergoes a Ne\"imark--Hopf bifurcation and becomes repulsive when giving rise to an attractive $ICC$ $(C)$ (Figure \ref{fig18}). Oscillations appear on $(C)$ when it is tangent to the cusp point $Q_c$ on $LC_b$ (Figure \ref{fig19}). From $\lambda=1.11$, $\lambda$ increasing, there is an alternation between existence of $ICC$ with oscillations, frequency lockings, cyclic chaotic behaviors, contact bifurcations with basin boundaries and single chaotic attractor. These results are consistent with the Liapunov exponents values obtained in \cite{Lo1}. We are going to explain what occurs regarding contact with basins of coexisting period--3 attractors.
\end{itemize}

\begin{itemize}
	\item In the same time, two attractive period--3 orbits appear simultaneously by a fold bifurcation, coexisting with two period--3 orbits of saddle points (Figure \ref{fig19}). So, three different attractors coexist; Figure \ref{fig20} shows the basin of the $ICC$ and the basin of the two period--3 attractors. When $\lambda$ increases, the two period--3 attractive cycles undergo a Ne\"imark--Hopf bifurcation and become unstable by giving rise to two period--3 attractive $ICC$ around them. Figure \ref{fig21} shows the three different basins of the three coexisting attractors : in this case, two period--3 $ICC$ and a period--29 attractive cycle, which corresponds to a frequency locking for the $ICC$. 
\end{itemize}

\begin{itemize}
	\item When $\lambda$ increases, the $ICC$ appears again; curles arise, due to crossings of $(C)$ through critical curves, and involve auto--intersections of $(C)$, the $ICC$ is then called a $weakly\ chaotic\ ring\ (WCR)$ \cite{Mi2} (Figure \ref{fig23}). The existence of curles changes the shape of the $ICC$ and generates contact with its basin boundary; this implies the explosion of $(C)$ and makes it disappear. Then, the basin of the former $ICC$ becomes part of the basin of the other attractors (Figures \ref{fig26} $\&$ \ref{fig28}).
\end{itemize}

\begin{itemize}
	\item Figure \ref{fig27} shows the separate basins of each part of the two period--3 attractive $ICC$, that means basins for $T^3$ are plotted. 
\end{itemize}

\begin{itemize}
	\item Another contact bifurcation occurs, which involves the manifolds of the period--3 saddle cycles, born by fold bifurcation. This contact creates a new attractor, by mixing the two former period--3 $ICC$ (Figure \ref{fig29}). This new attractor evolves, undergoes frequency lockings, which give rise to the appearance of cyclic chaotic attractor (Figure \ref{fig30}); curles appear on the attractor and it becomes a $WCR$ (Figures \ref{fig36}--\ref{fig38}), then it becomes a chaotic attractor; chaos becomes stronger and stronger (Figures \ref{fig43}--\ref{fig44}).
\end{itemize}

\begin{itemize}
	\item The chaotic attractor disappears when a contact bifurcation occurs between the attractor and its basin boundary (Figure \ref{fig46}).
\end{itemize}

\section{Conclusion}
Complex behaviour of a two-dimensional noninvertible map with a multiplicative global coupling between one-dimensional logistic maps, including a time asymmetric feedback, has been analyzed. The different kinds of basin bifurcations which occur when the parameter is modified and the evolution of attractors issued from Ne\"imark--Hopf bifurcation have been explained. The fundamental point, as previously obtained in two-dimensional noninvertible maps studies, is the position of basins and attractors, regarding critical curves. $Islands\ aggregations$, $connected\ basin\ \leftrightarrow\ nonconnected$ $basin$, $connected\ basin$ $\leftrightarrow$ $multiply-connected\ basin$, change of $Invariant\ Closed\ Curve$ into $Weakly\ Chaotic\ Ring$ or $Strong\ Chaotic\ Attractor$ are bifurcations in relation with critical curves.

\section{Acknowledgment}
\begin{figure}[t]
   \begin{minipage}[c]{.50\linewidth}
      \includegraphics[width=0.9\textwidth]{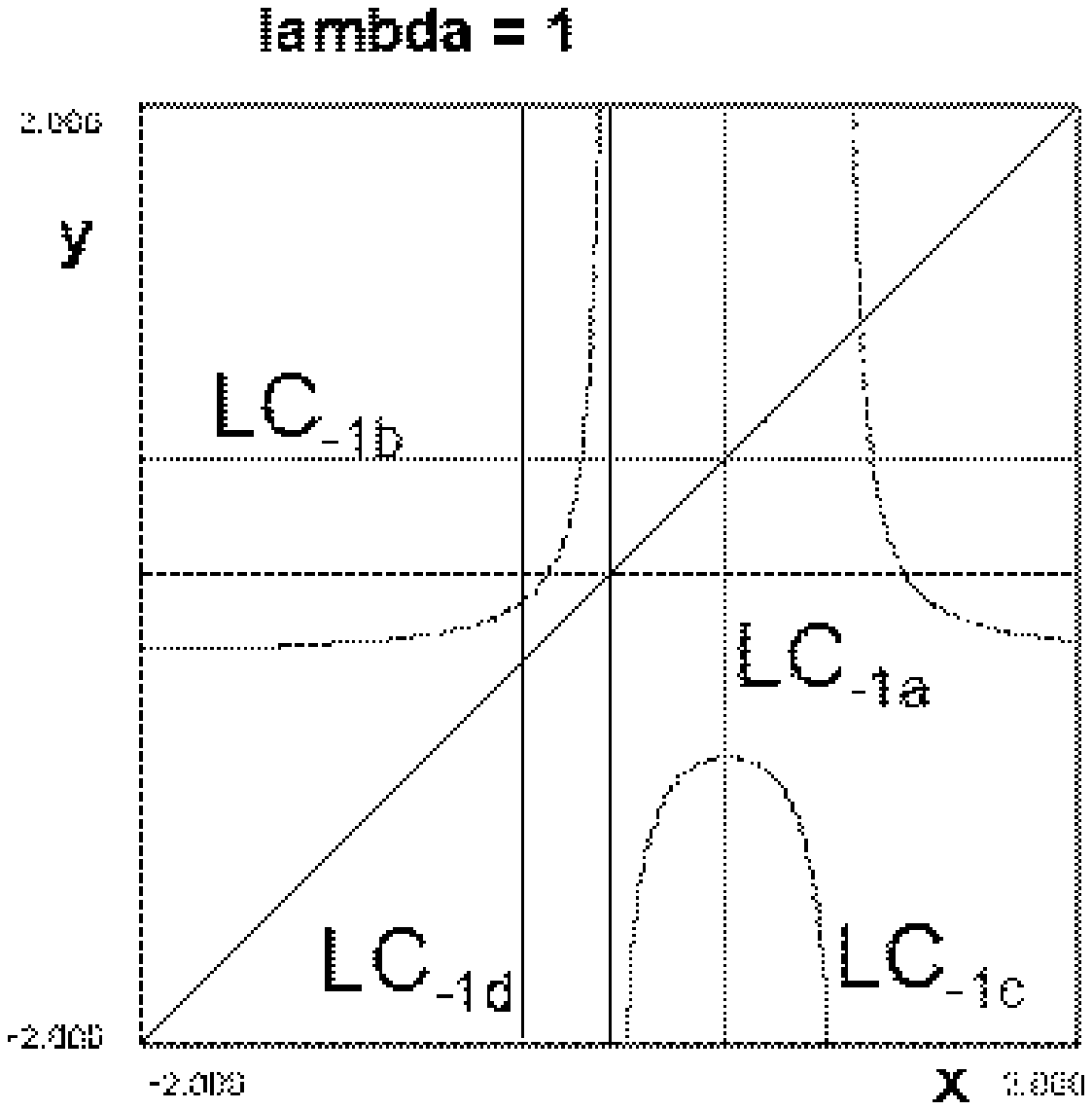}
      \caption[width=0.80]{Preimages of critical lines:\\ $LC_-$$_1$ arcs.\label{fig1}}
   \end{minipage} \hfill
   \begin{minipage}[c]{.50\linewidth}
      \includegraphics[width=0.80\textwidth]{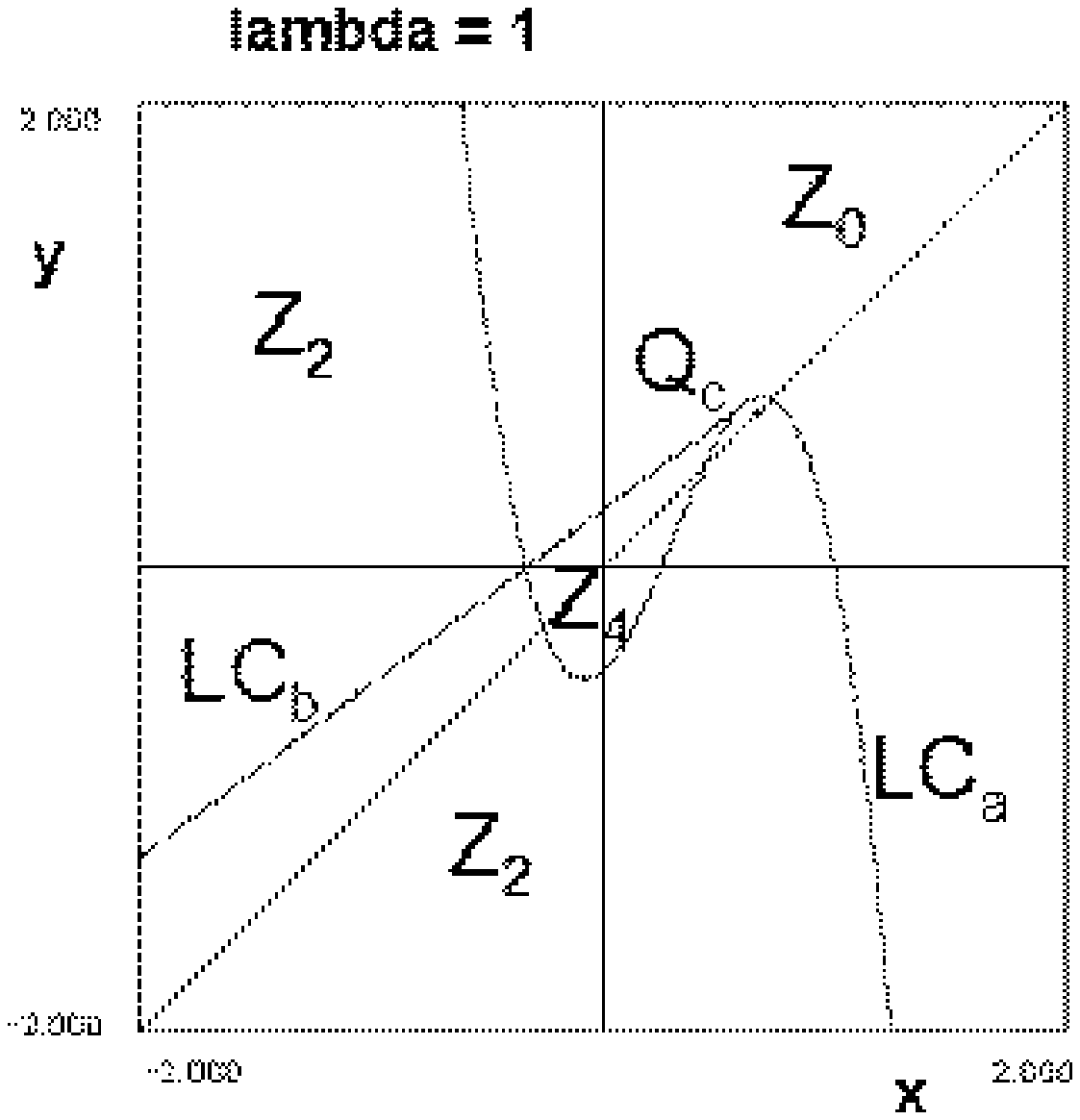}
      \caption[width=0.80]{Critical lines LC and $Z_i$ areas in the state plane.\label{fig2}}
   \end{minipage}
	\begin{center}
		\includegraphics[width=0.50\textwidth]{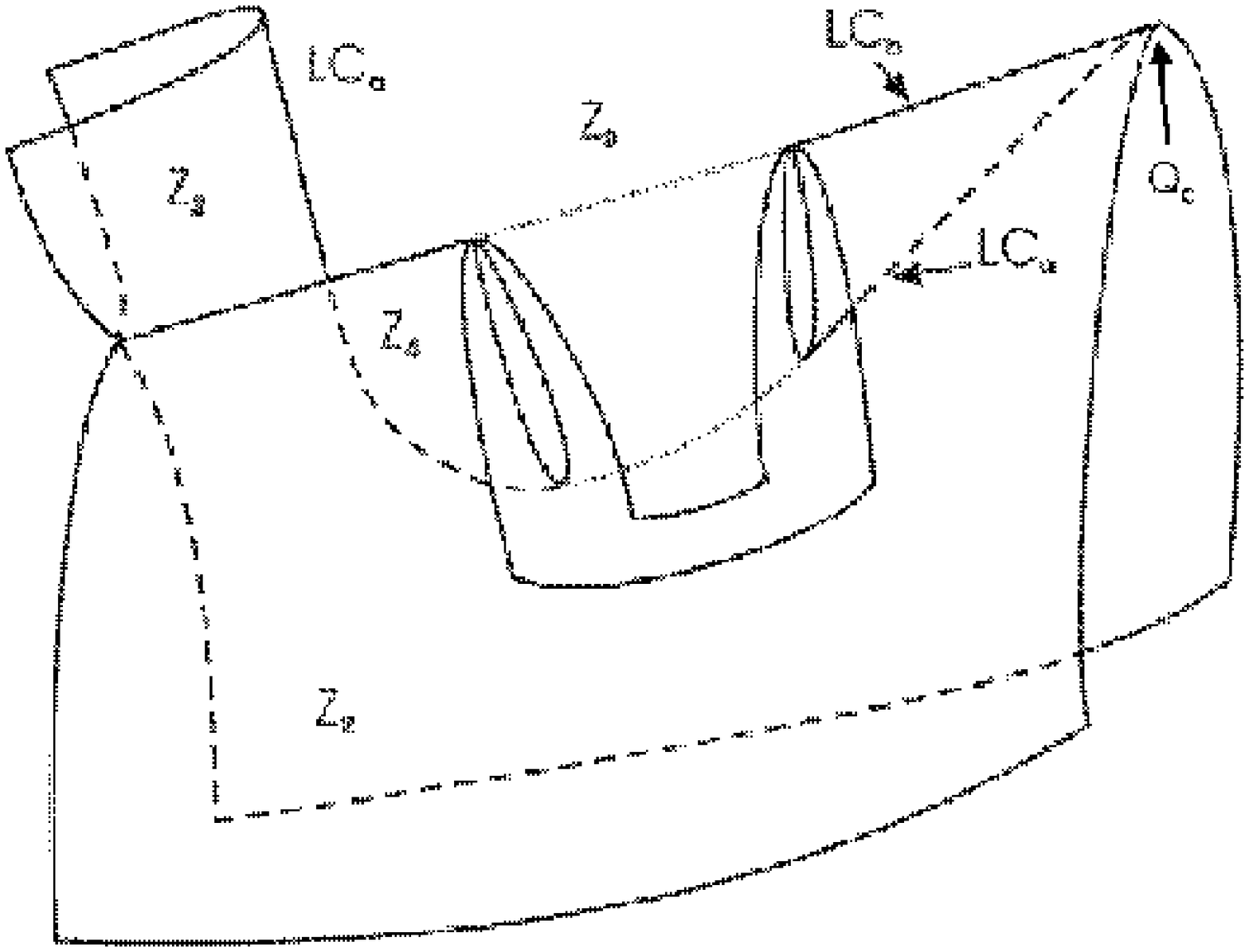}
	\end{center}
	\caption[width=0.80]{Qualitative representation of the foliation of the state space, the open U-shaped window permits to show a section of $Z_4$ and $Z_2$ area. On the right end, one can see the cusp point $Q_c$ on $LC_b$.}\label{fig2a}
\end{figure}

\begin{figure}[t]
   \begin{minipage}[c]{.50\linewidth}
      \includegraphics[width=0.80\textwidth]{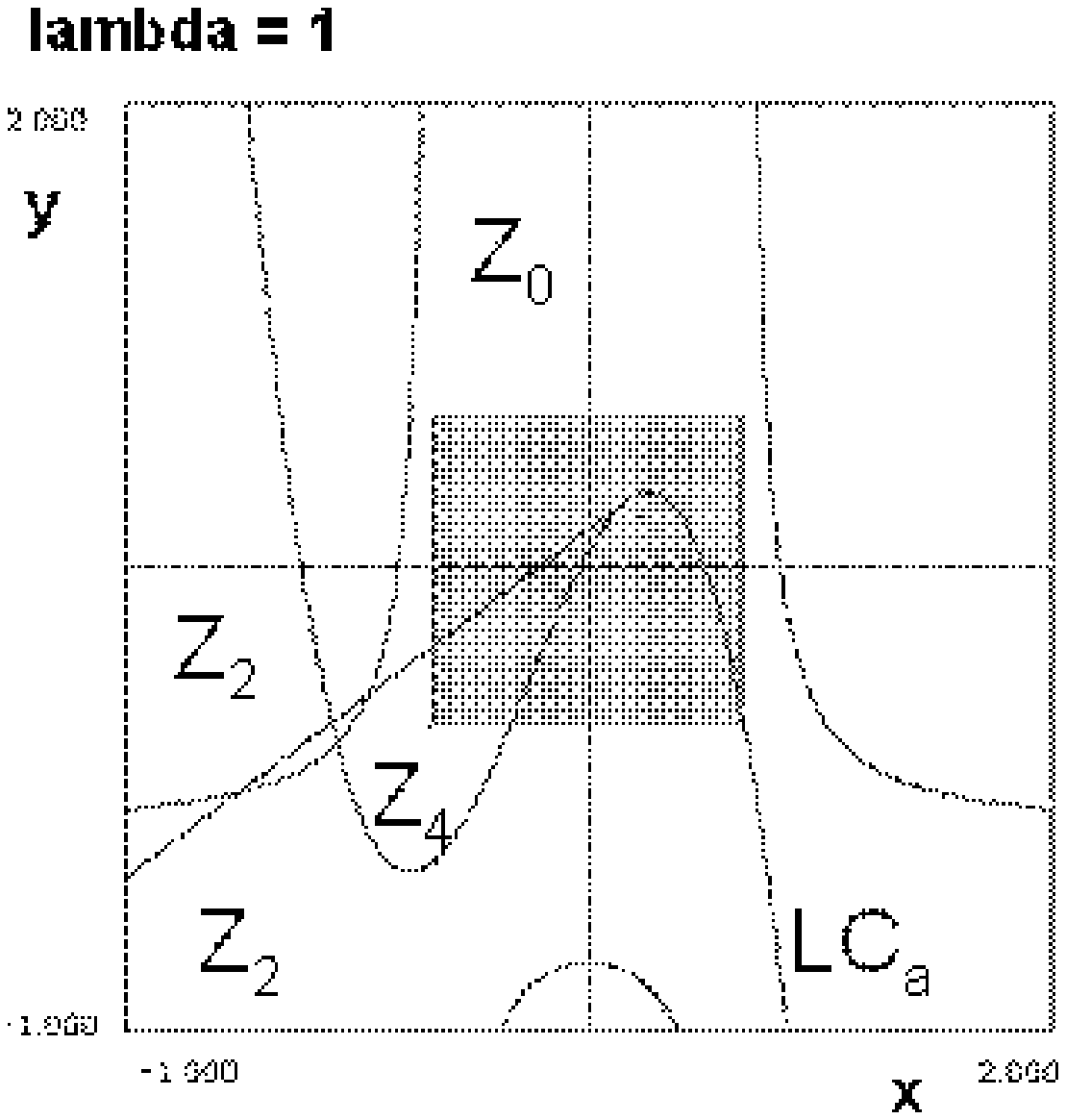}
      \caption[width=0.80]{$\lambda$=1, the basin is a square. \label{fig3}}
   \end{minipage} \hfill
   \begin{minipage}[c]{.50\linewidth}
      \includegraphics[width=0.80\textwidth]{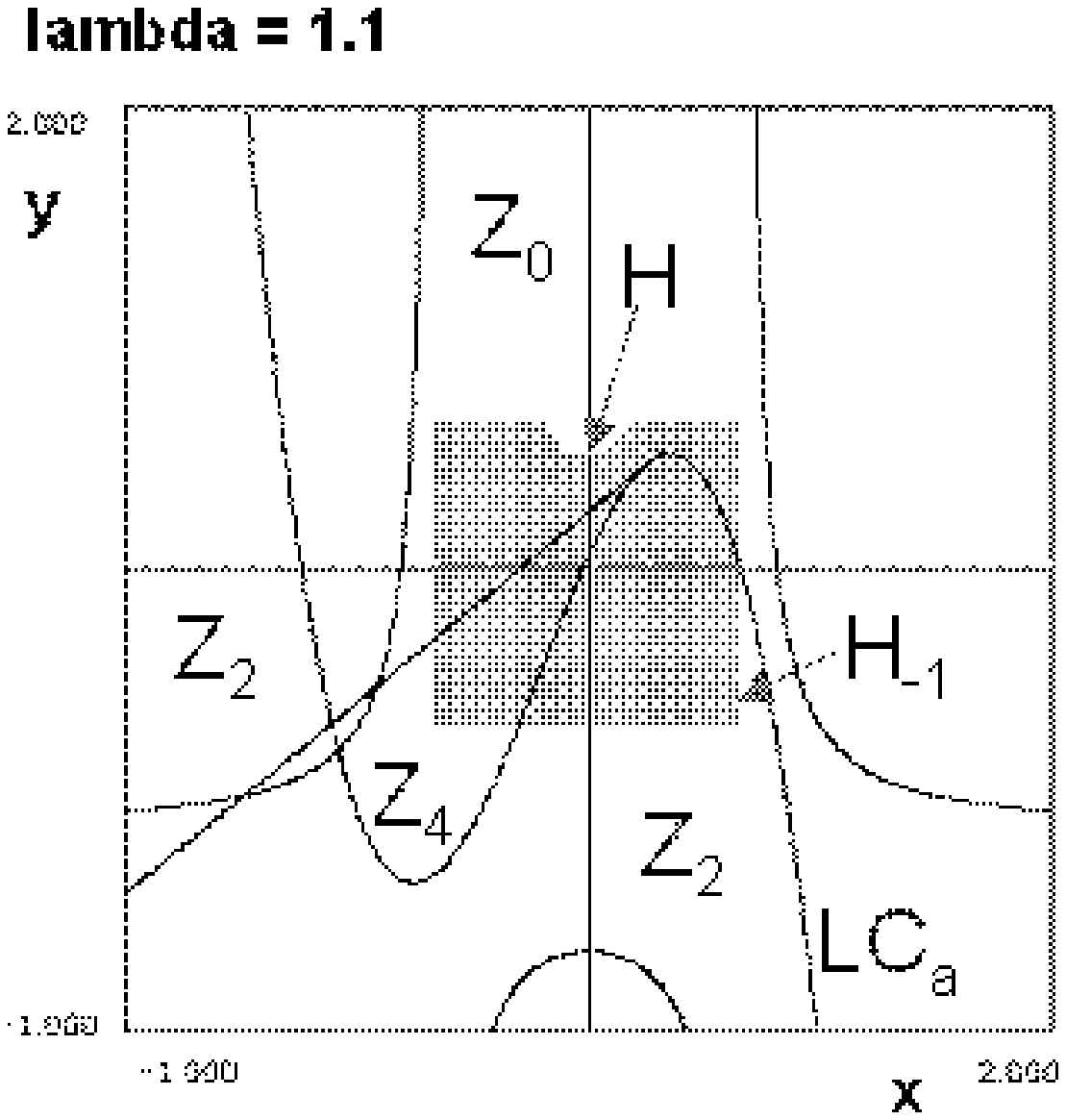}

      \caption[width=0.80]{Creation of a bay H when $LC_a$ crosses through 
      basin boundary; $\lambda$$\geq$1 \label{fig4}}.
   \end{minipage}
	\begin{center}
		\includegraphics[width=0.50\textwidth]{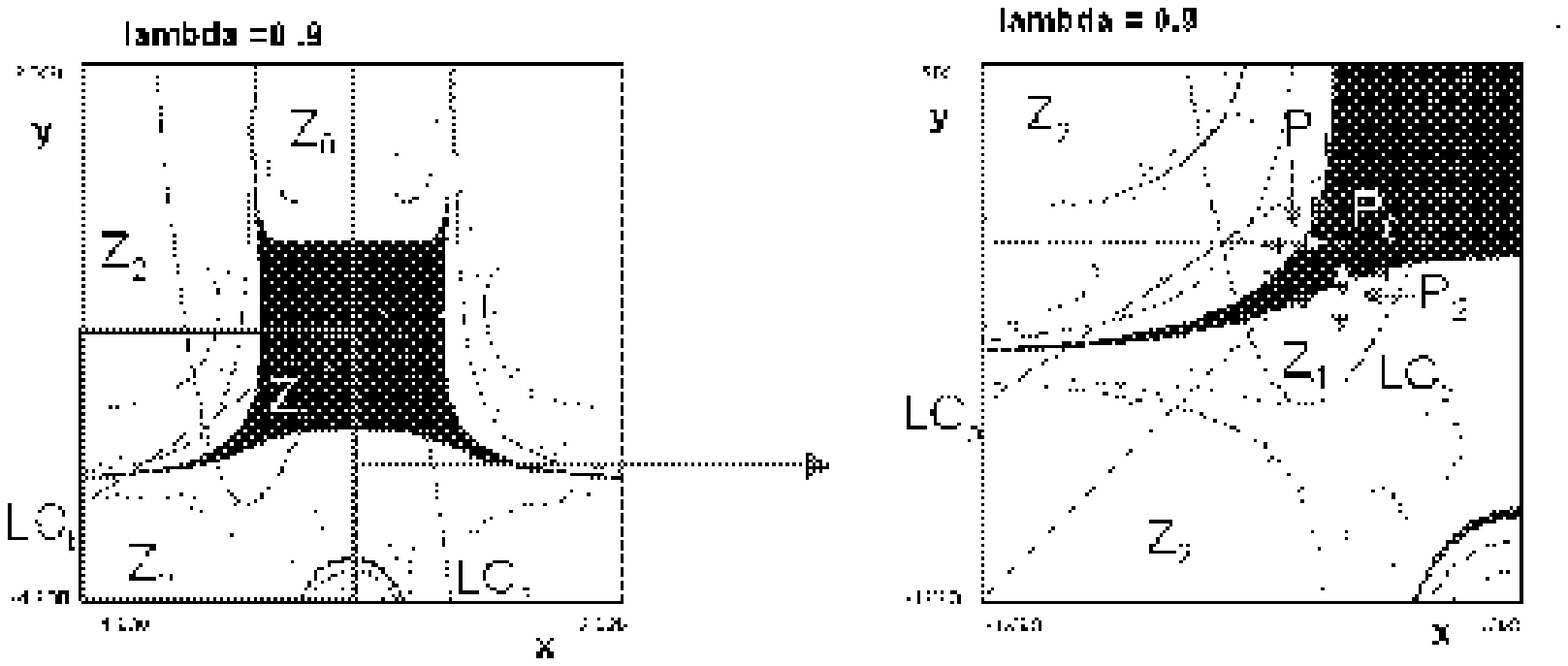}
	\end{center}
	\caption[width=0.90]{Creation of islands.}
	\label{fig6}
\end{figure}

\begin{figure}[t]
   \begin{minipage}[c]{.50\linewidth}
      \includegraphics[width=0.80\textwidth]{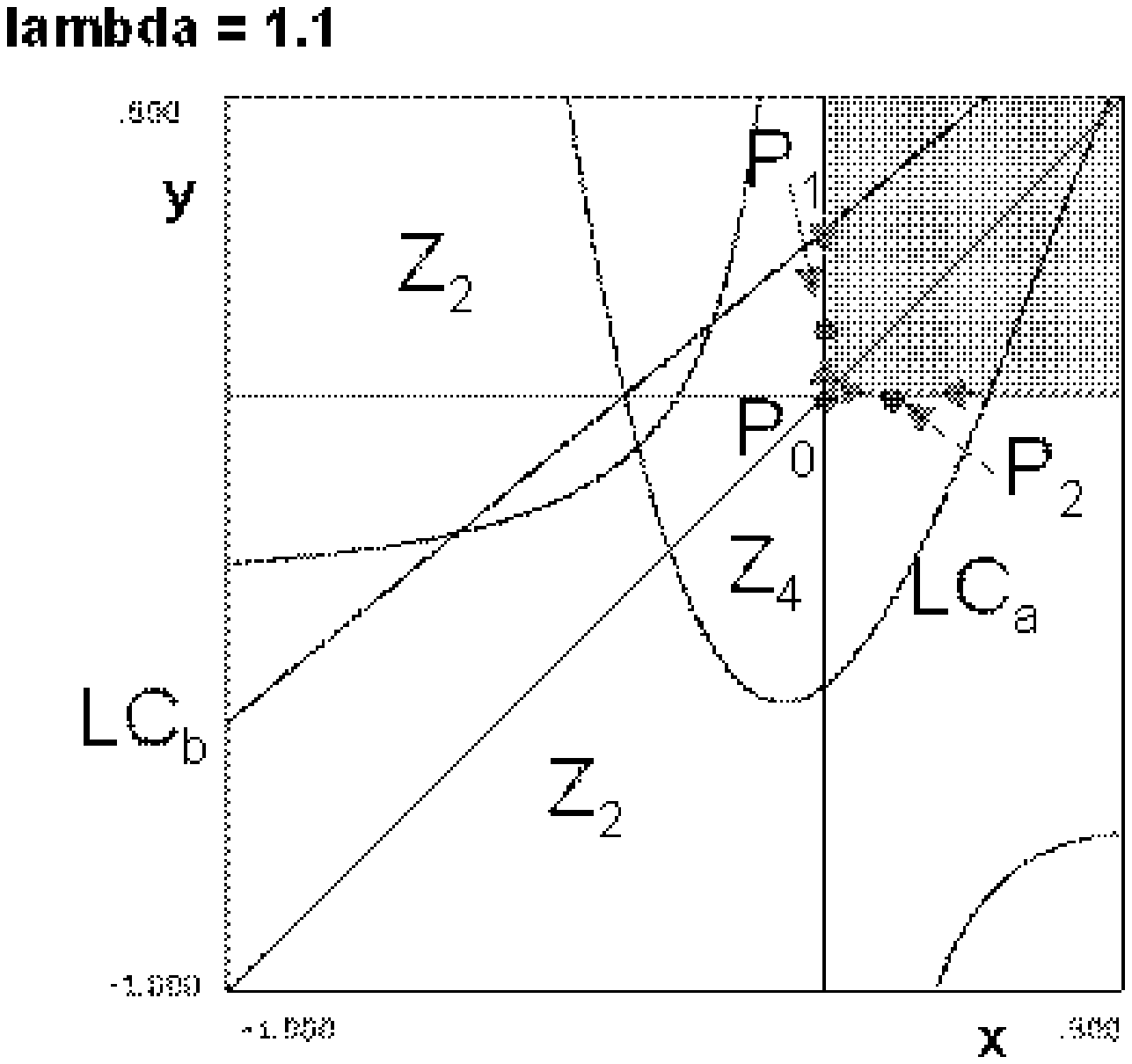}
      \caption[width=0.80]{Before the bifurcation\\ connected basin $\leftrightarrow$
      nonconnected basin. \label{fig7}}
   \end{minipage} \hfill
   \begin{minipage}[c]{.50\linewidth}
      \includegraphics[width=0.80\textwidth]{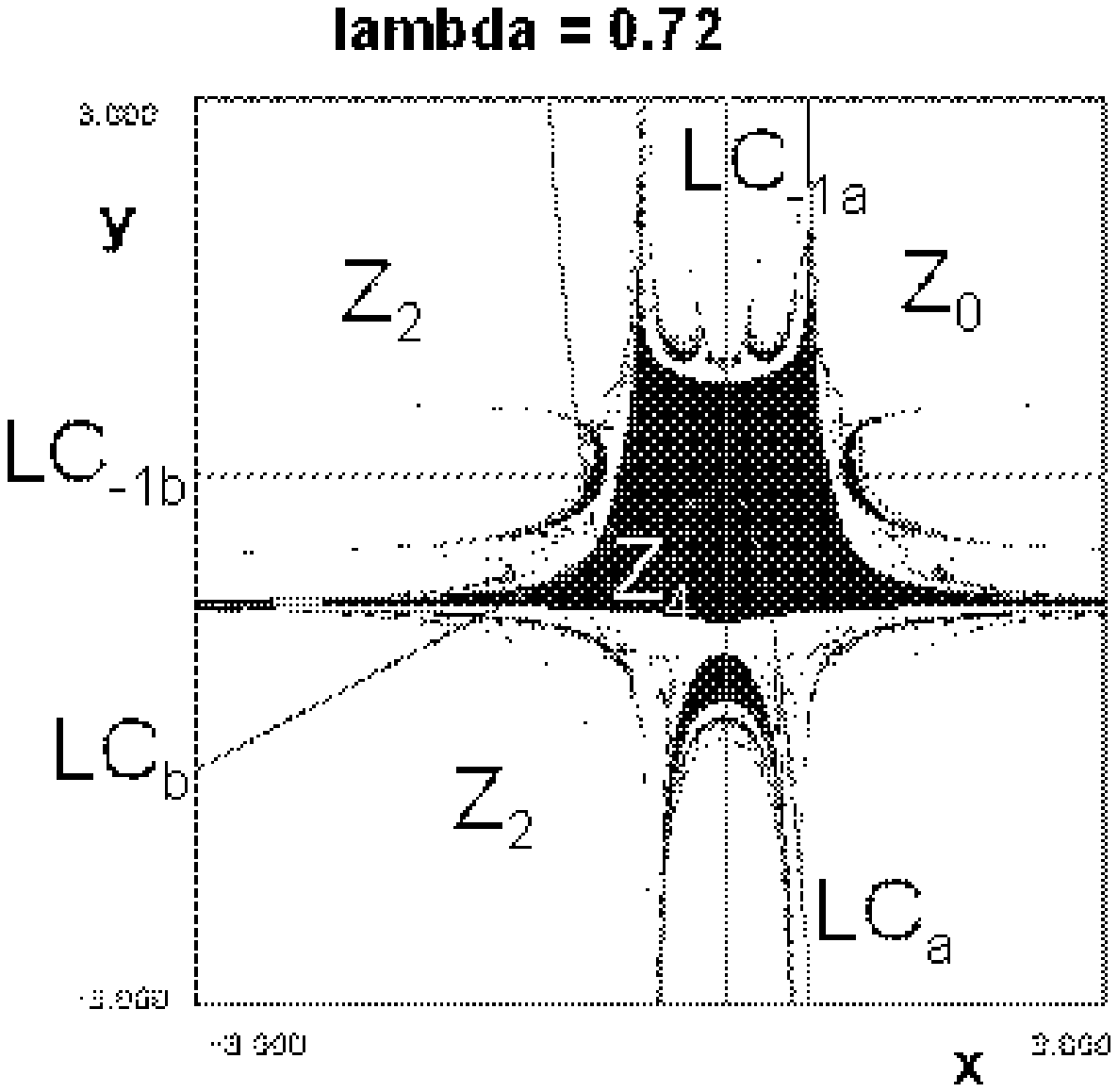}
      \caption[width=0.80]{Before islands aggregation.\label{fig9}}
   \end{minipage}
\end{figure}

\begin{figure}[t]
   \begin{minipage}[c]{.50\linewidth}
      \includegraphics[width=0.80\textwidth]{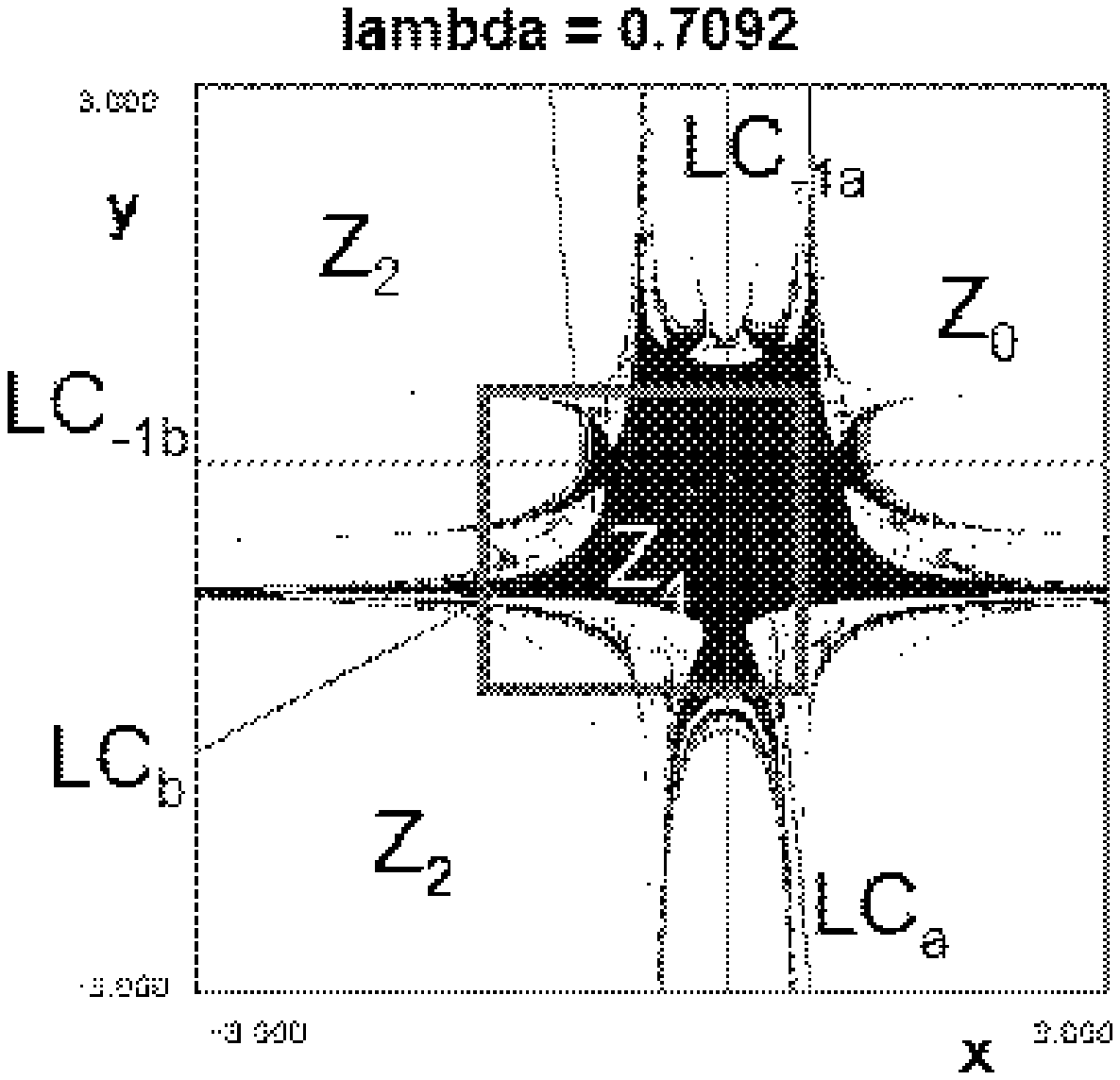}
      \caption[width=0.80]{Islands aggregation. \label{fig10}}
   \end{minipage} \hfill
   \begin{minipage}[c]{.50\linewidth}
      \includegraphics[width=0.80\textwidth]{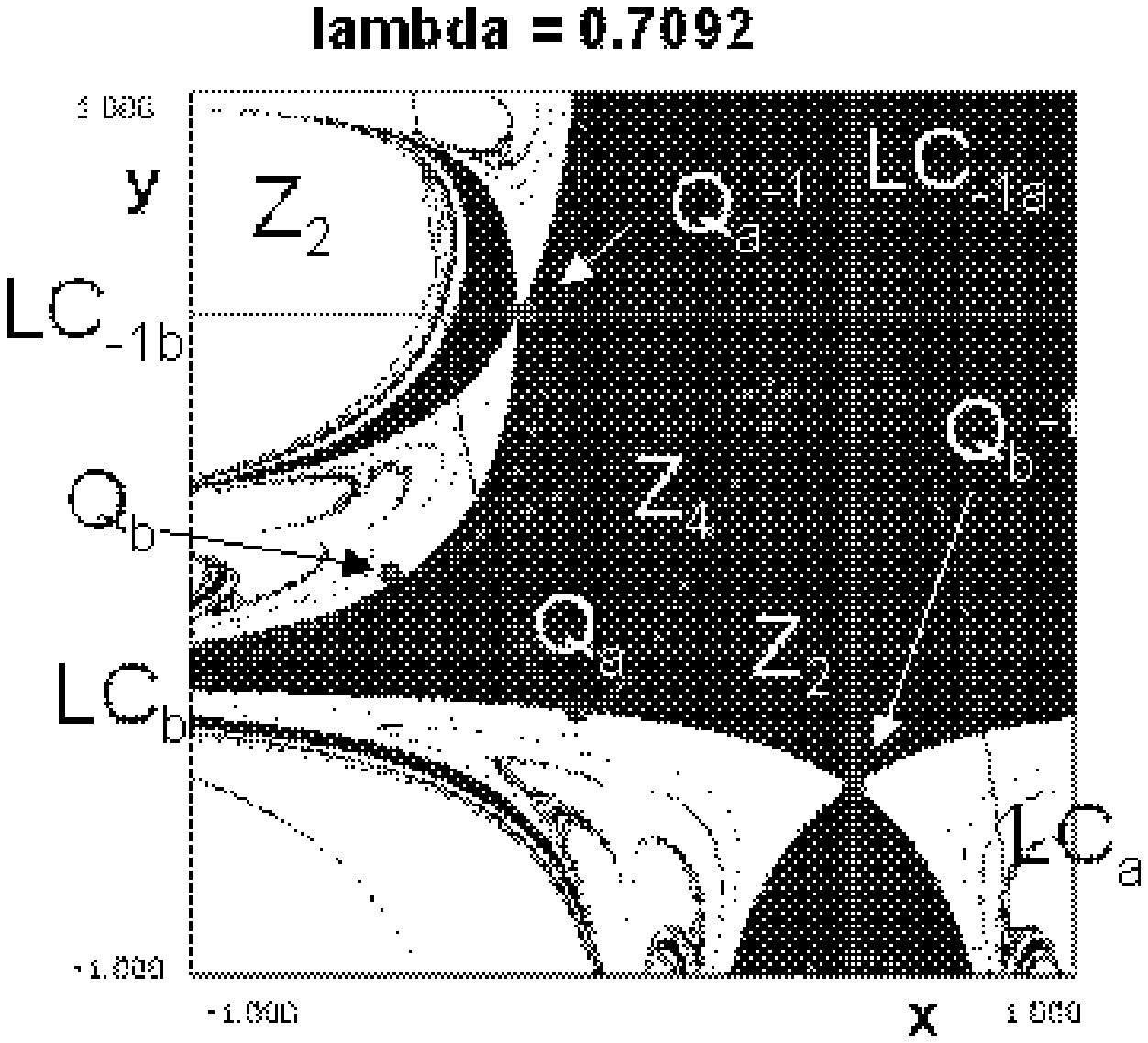}
      \caption[width=0.80]{Enlargment of Figure \ref{fig10}. \label{fig9a}}
   \end{minipage}
\end{figure}

\begin{figure}[t]
   \begin{minipage}[c]{.50\linewidth}
      \includegraphics[width=0.80\textwidth]{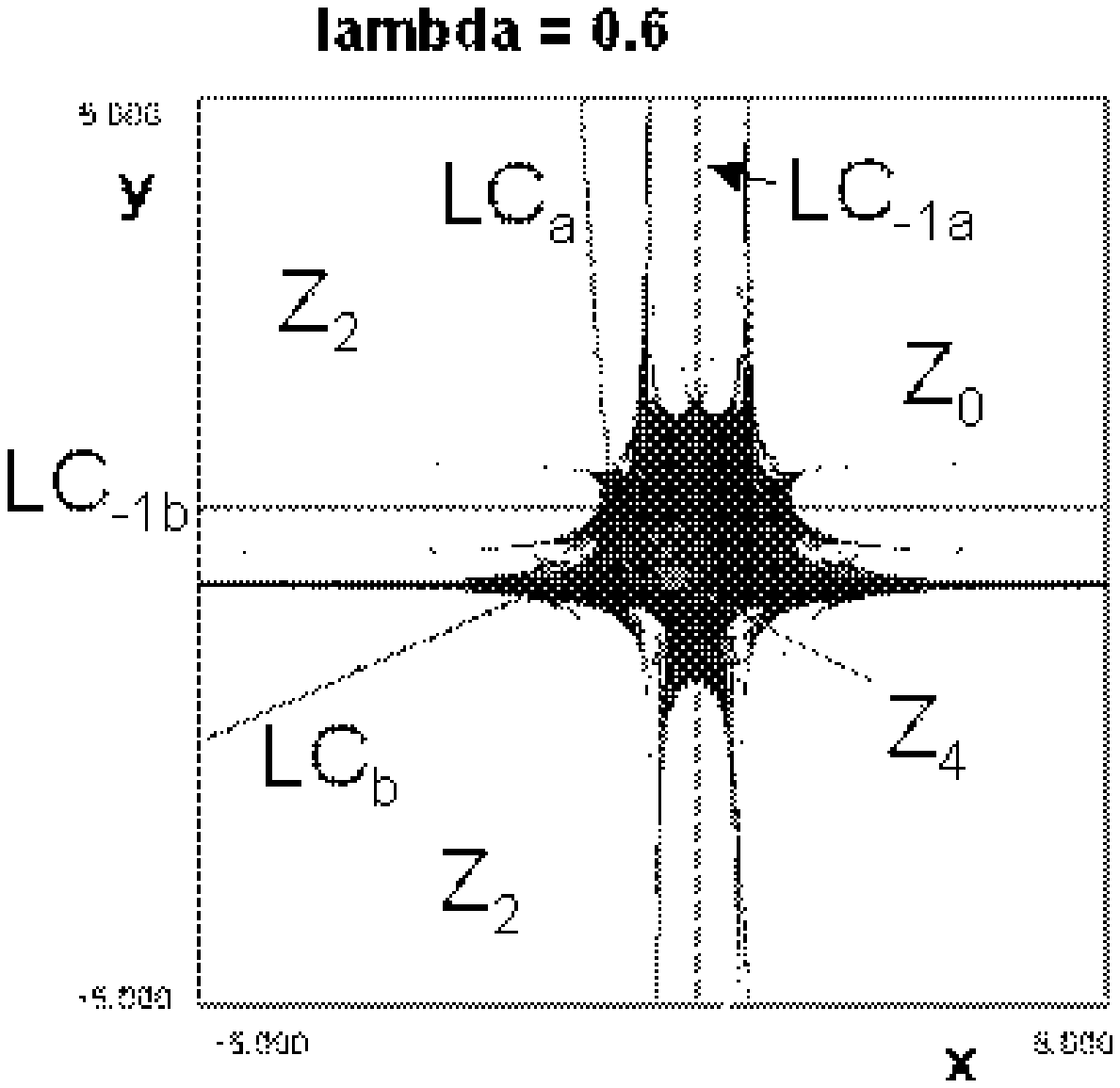}
      \caption[width=0.80]{A sequence of islands\\ aggregations occurs when $\lambda$ decreases. \label{fig11}}
   \end{minipage} \hfill
   \begin{minipage}[c]{.50\linewidth}
      \includegraphics[width=0.80\textwidth]{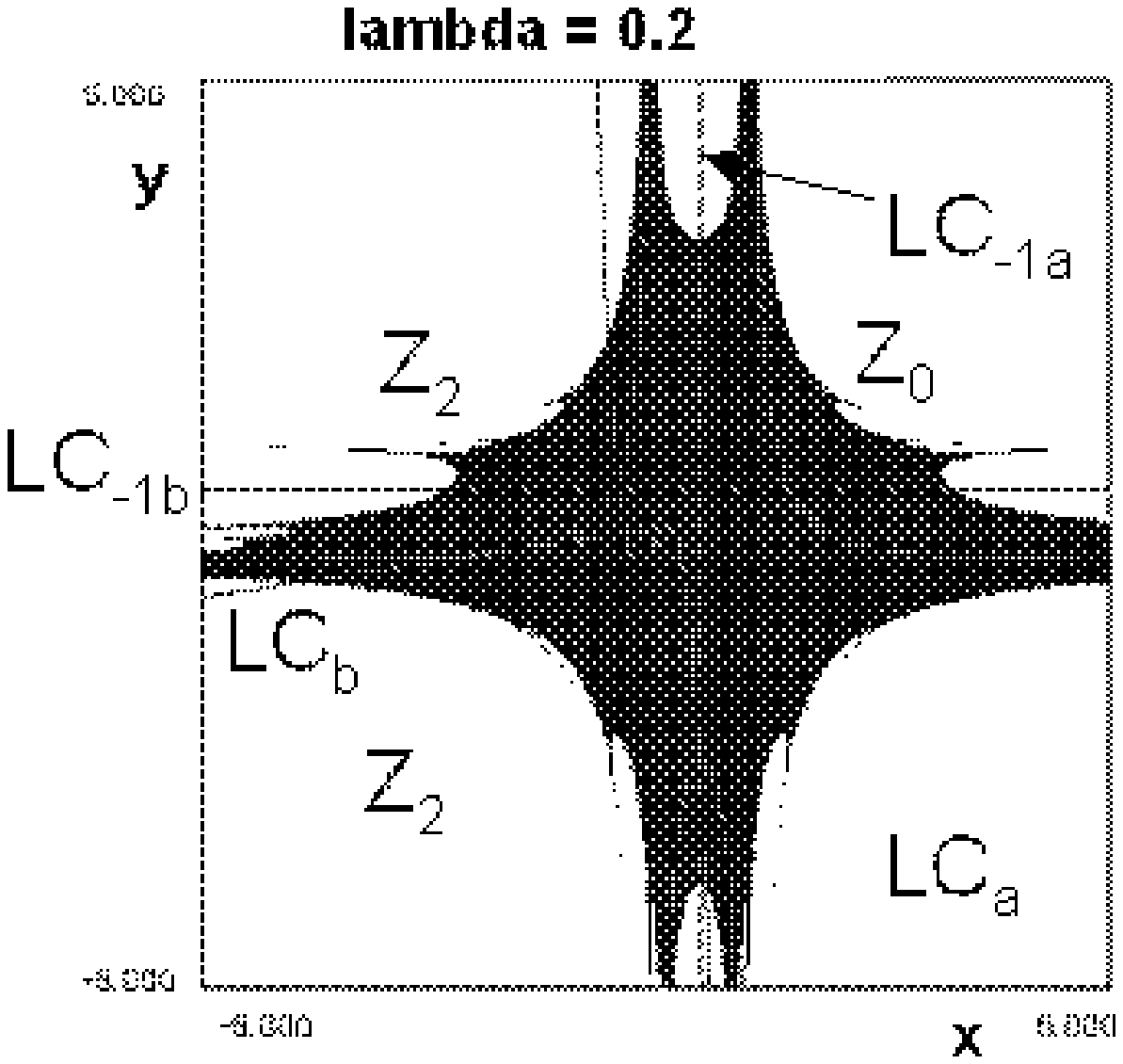}
      \caption[width=0.80]{Tongues disappear by crossing of basin boundary through critical curves.\label{fig12}}
   \end{minipage}
\end{figure}

\begin{figure}[t]
   \begin{minipage}[c]{.50\linewidth}
      \includegraphics[width=0.80\textwidth]{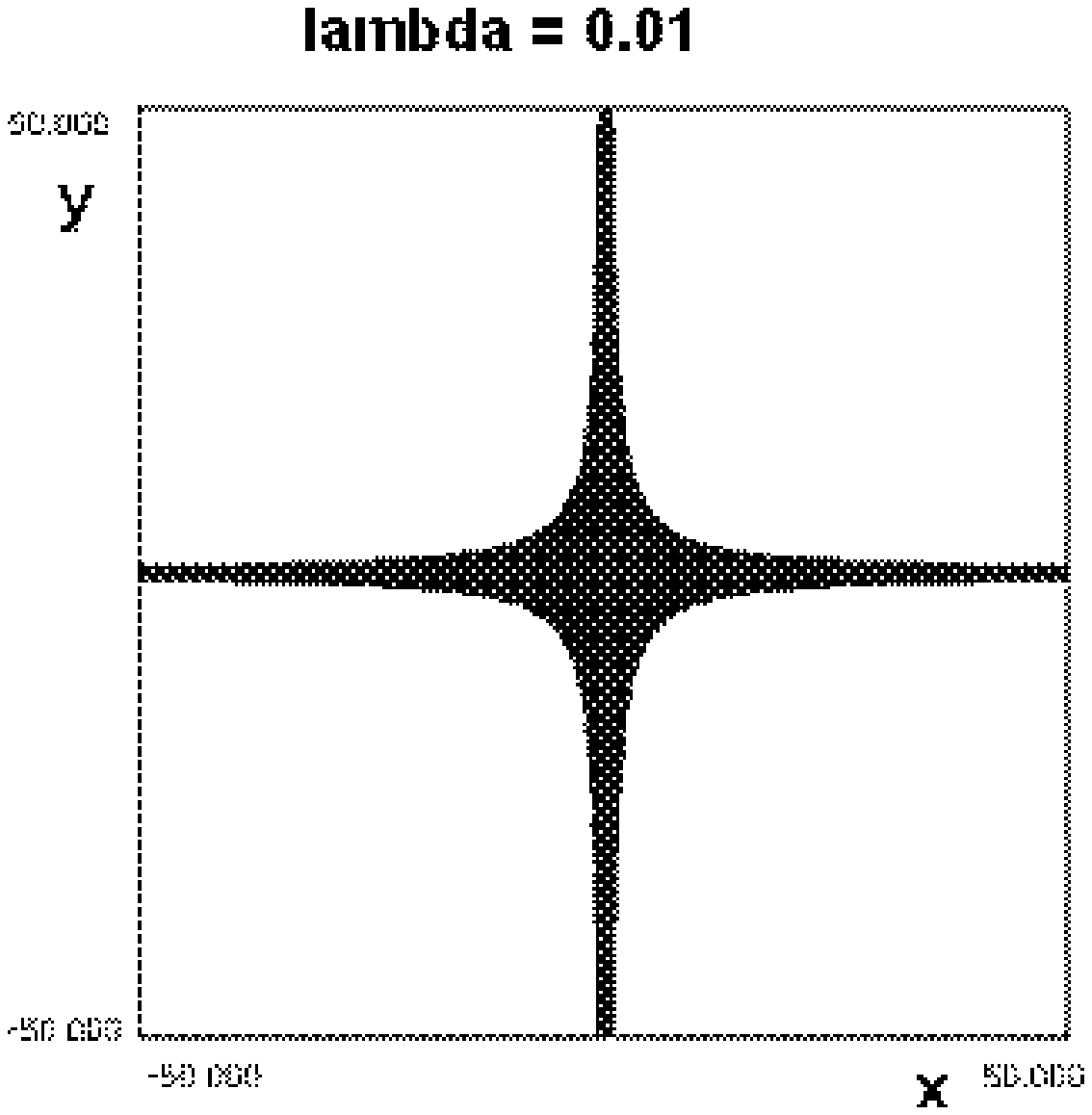}
      \caption[width=0.80]{The basin boundary is smooth. \label{fig13}}
   \end{minipage} \hfill
   \begin{minipage}[c]{.50\linewidth}
      \includegraphics[width=0.80\textwidth]{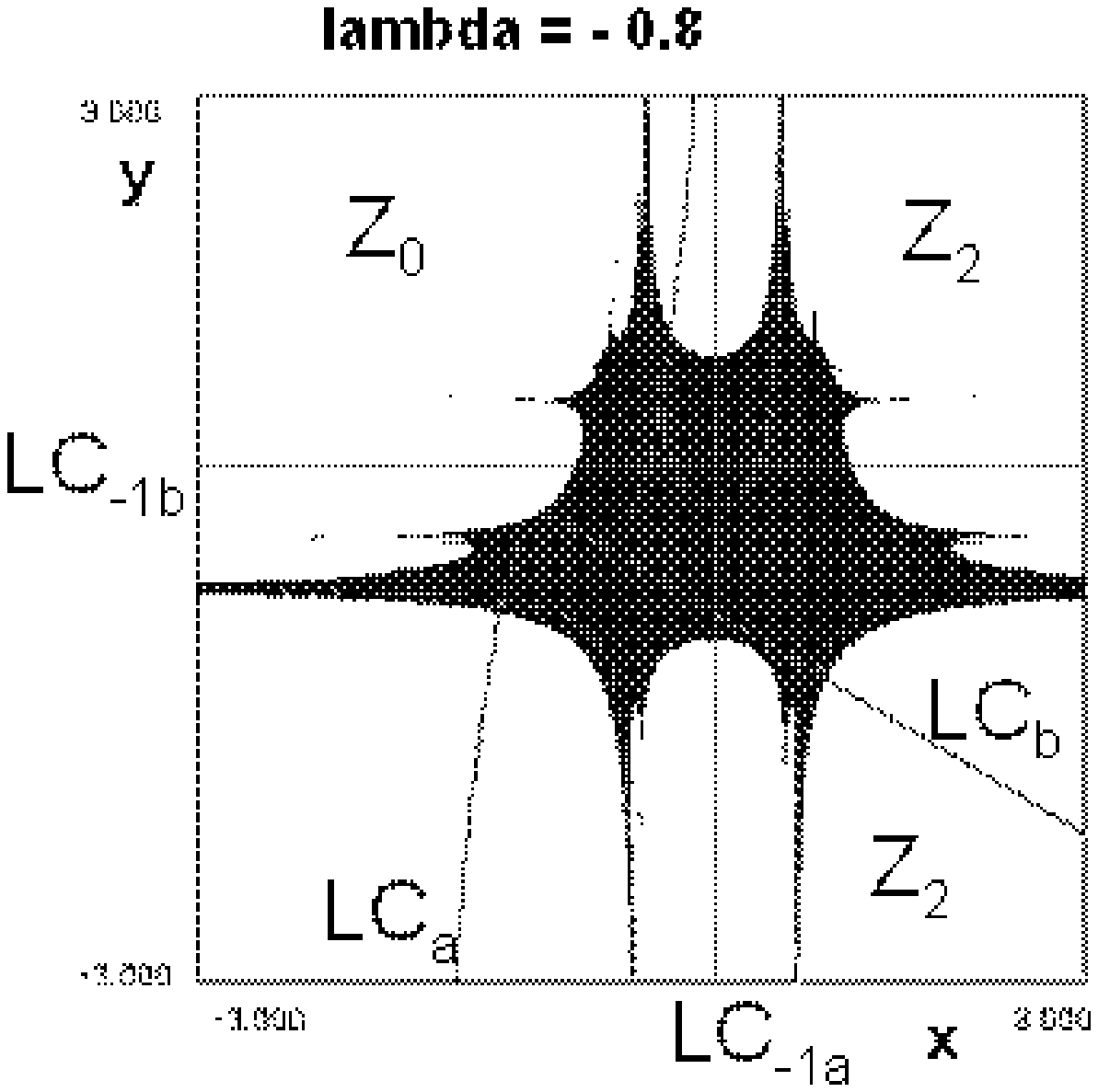}
      \caption[width=0.80]{Tongues appear again when $\lambda$$<$0. \label{fig14}}
   \end{minipage}
\end{figure}

\begin{figure}[t]
   \begin{minipage}[c]{.50\linewidth}
      \includegraphics[width=0.80\textwidth]{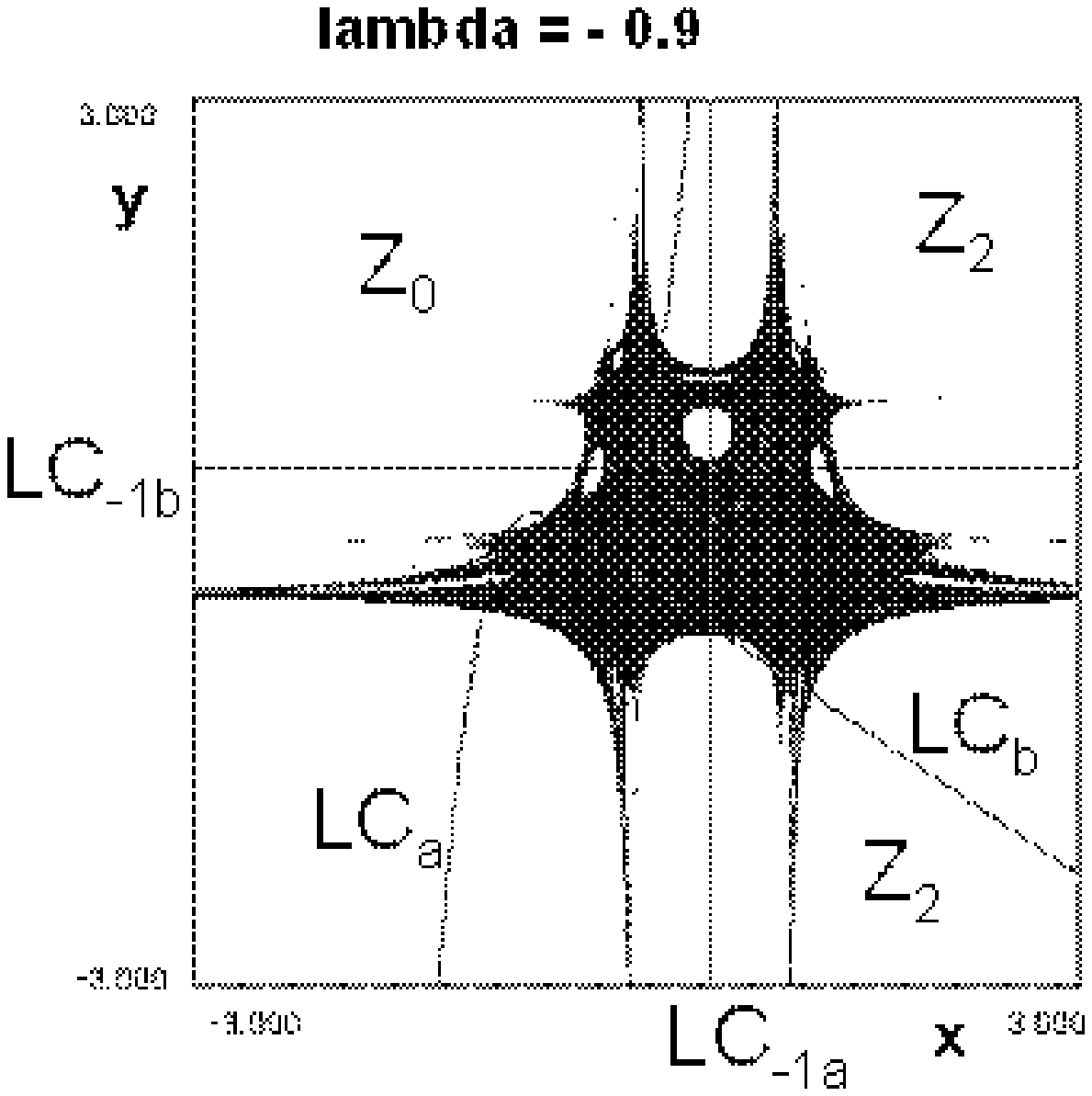}
      \caption[width=0.80]{Basin becomes multiply\\ connected. \label{fig15}}
   \end{minipage} \hfill
   \begin{minipage}[c]{.50\linewidth}
      \includegraphics[width=0.80\textwidth]{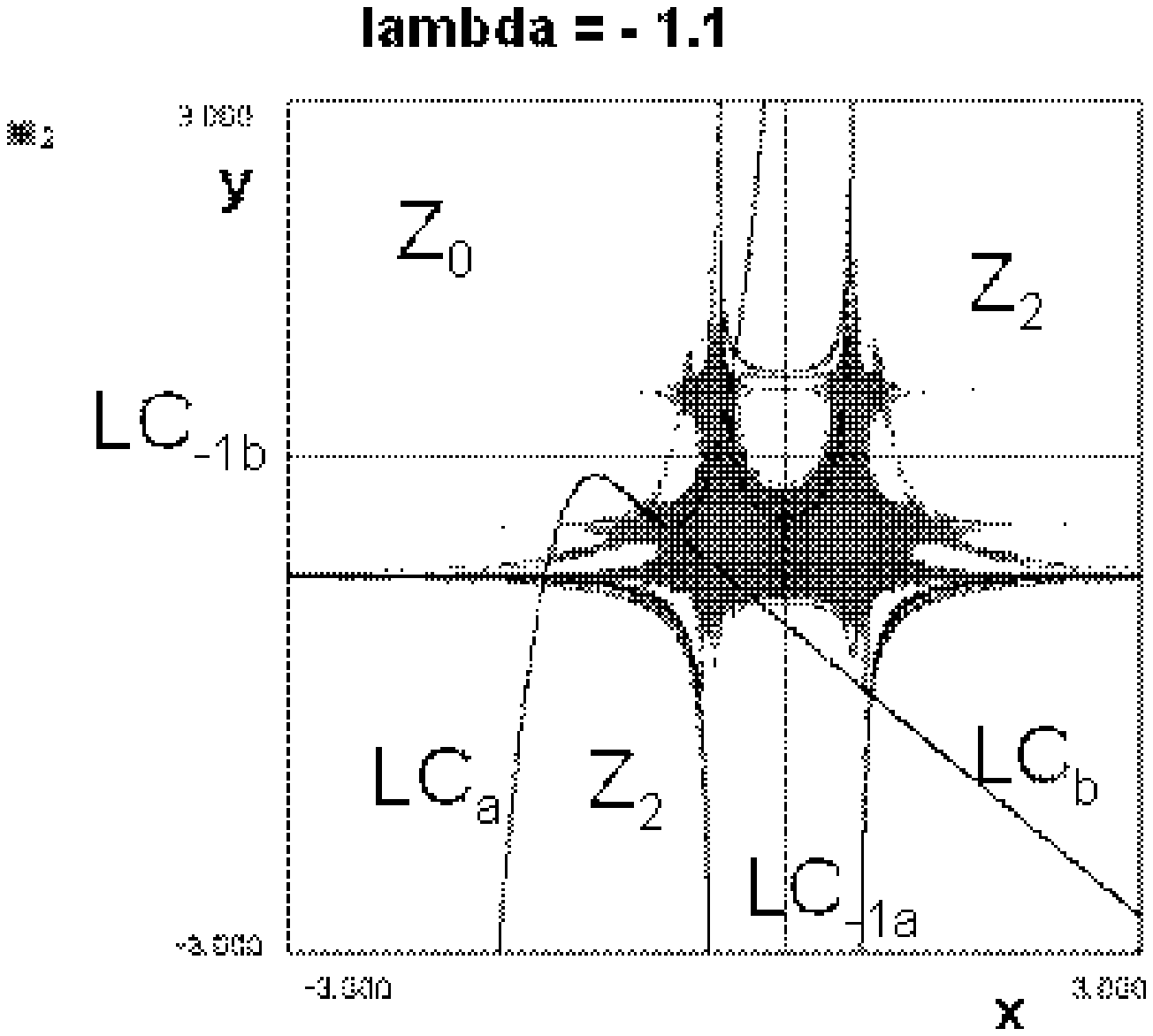}
      \caption[width=0.80]{When $\lambda$ decreases, the size of holes increases. $P_0$ has undergone a flip bifurcation: an attractive period--2 cycle exists.
       \label{fig16}}
   \end{minipage}
\end{figure}

\begin{figure}[t]
	\begin{center}
		\includegraphics[width=0.40\textwidth]{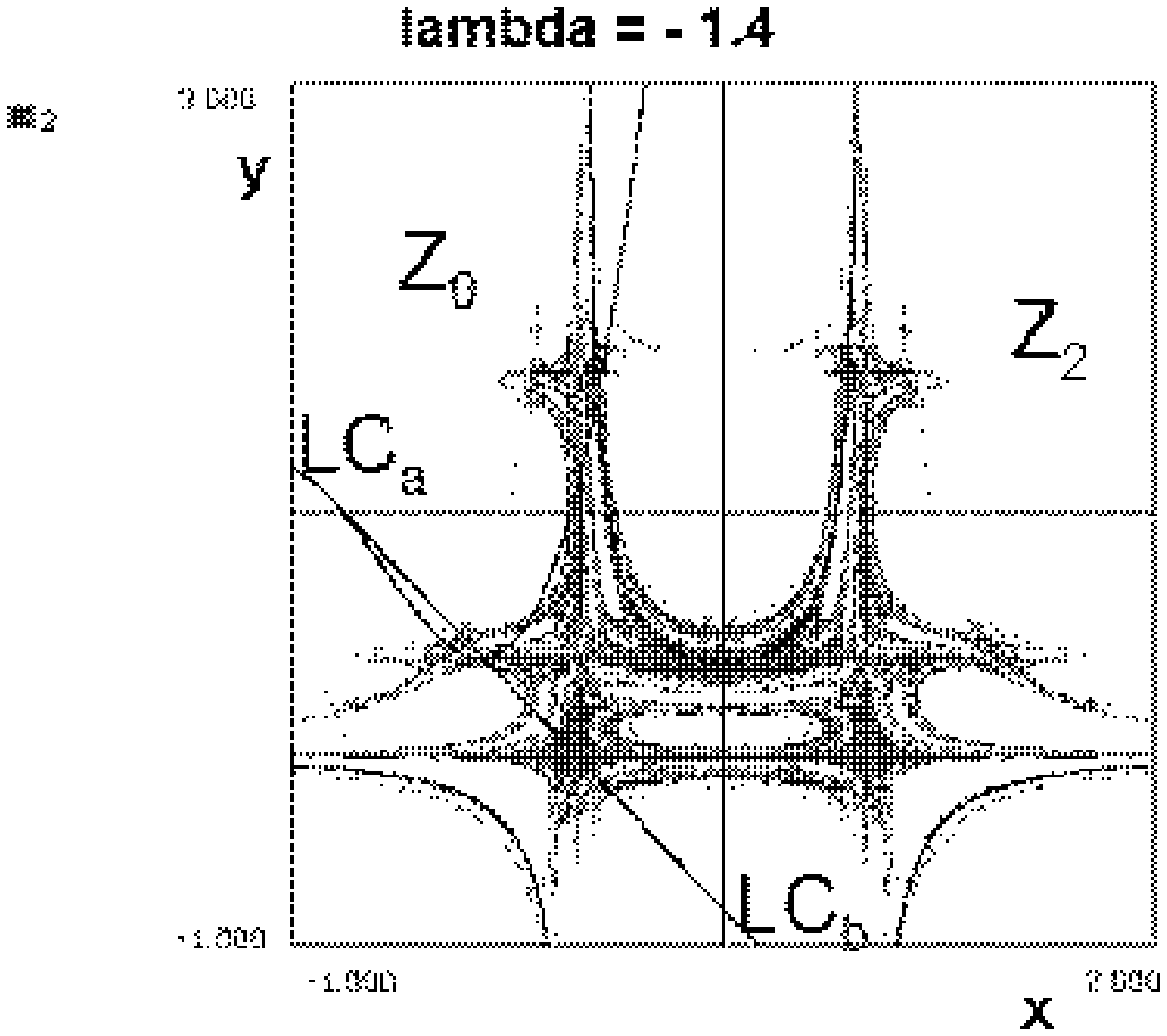}
	\end{center}
	\caption[width=0.80]{Basin boundary is fractal. An attractive period--2 cycle exists.}
	\label{fig17}
\end{figure}

\begin{figure}[t]
   \begin{minipage}[c]{.50\linewidth}
      \includegraphics[width=0.80\textwidth]{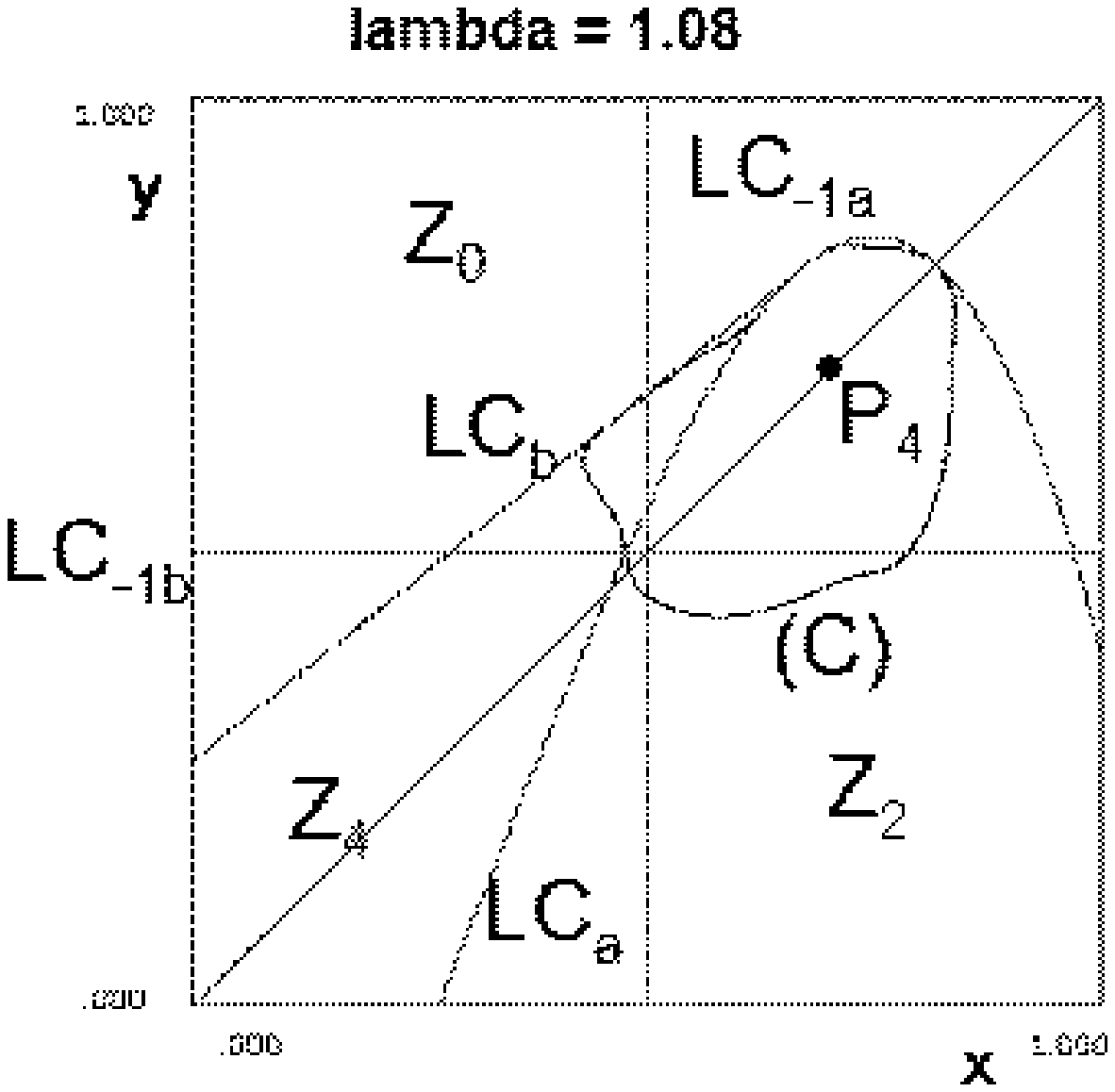}
      \caption[width=0.80]{An attractive invariant closed\\ curve $(C)$
       exists around the fixed point $P_4$. \label{fig18}}
   \end{minipage} \hfill
   \begin{minipage}[c]{.50\linewidth}
      \includegraphics[width=0.80\textwidth]{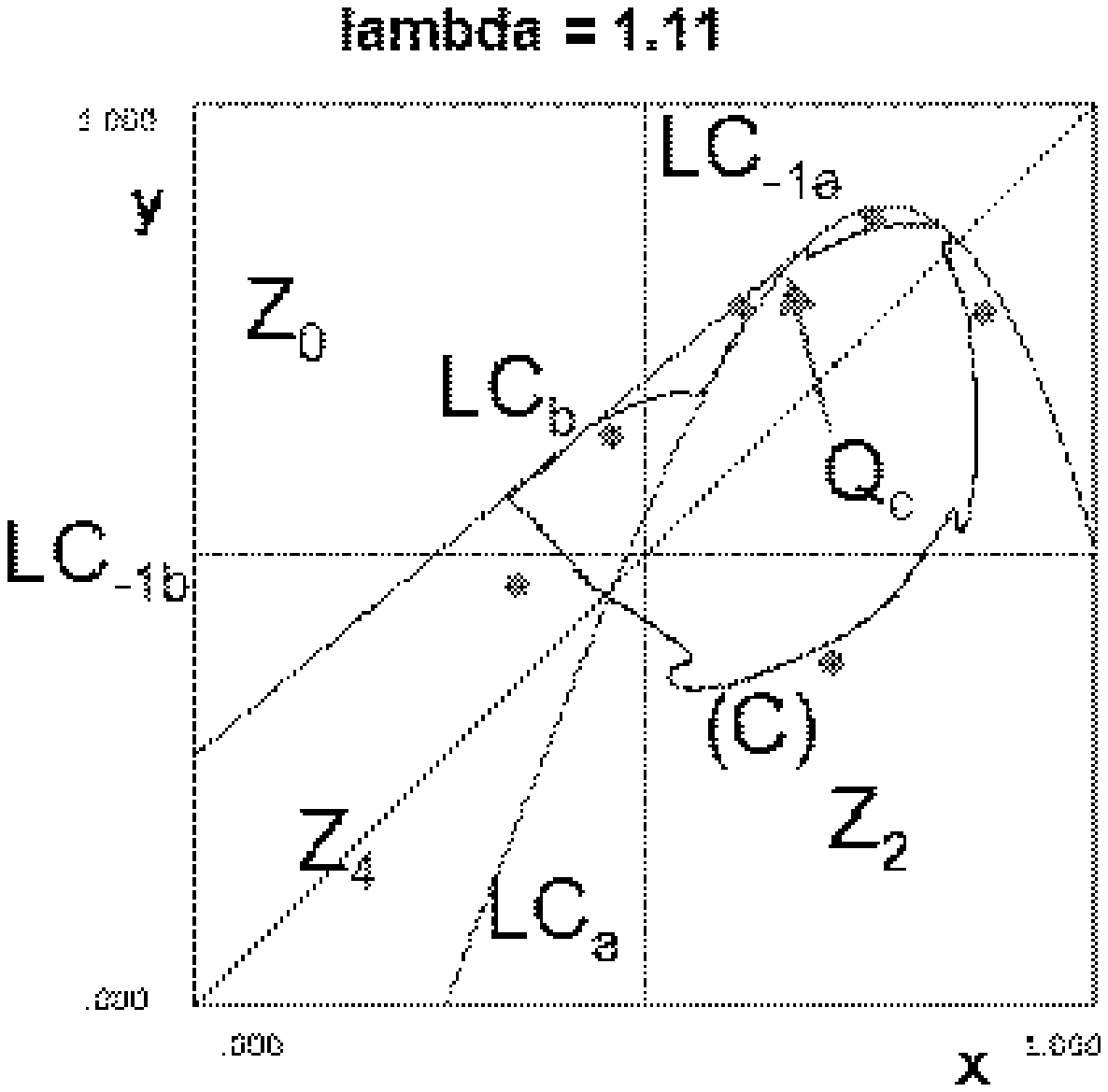}
      \caption[width=0.80]{Two stable 3-periodic orbits appear 
      by fold bifurcation; oscillations appear on $(C)$ when it is 
      tangent to the cusp point $Q_c$ on $LC_b$. \label{fig19}}
   \end{minipage}
   \begin{minipage}[c]{.50\linewidth}
      \includegraphics[width=0.80\textwidth]{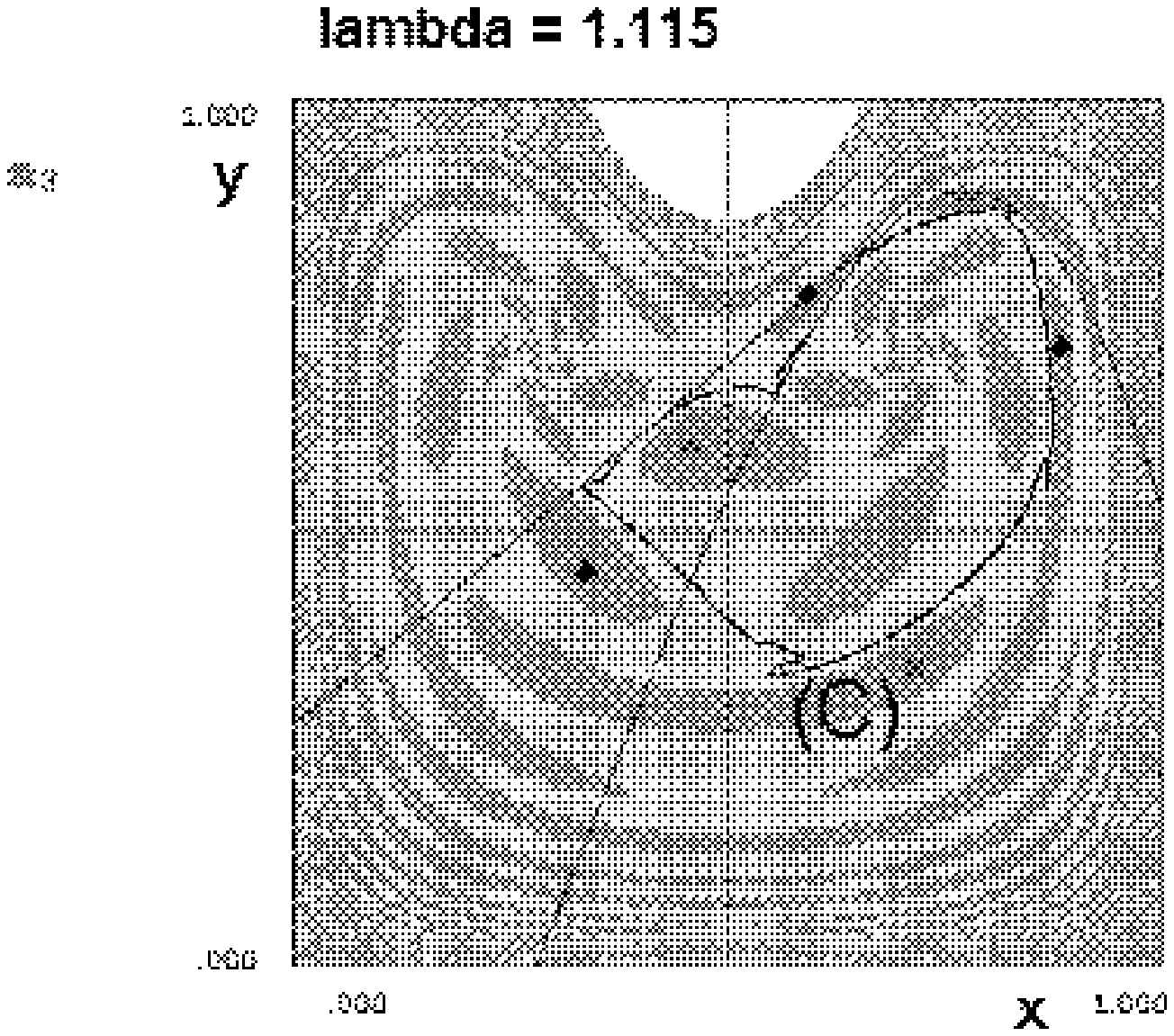}
      \caption[width=0.80]{Basins of the $ICC$ and of the\\ two period--3 attractive cycles. \label{fig20}}
   \end{minipage} \hfill
   \begin{minipage}[c]{.50\linewidth}
      \includegraphics[width=0.80\textwidth]{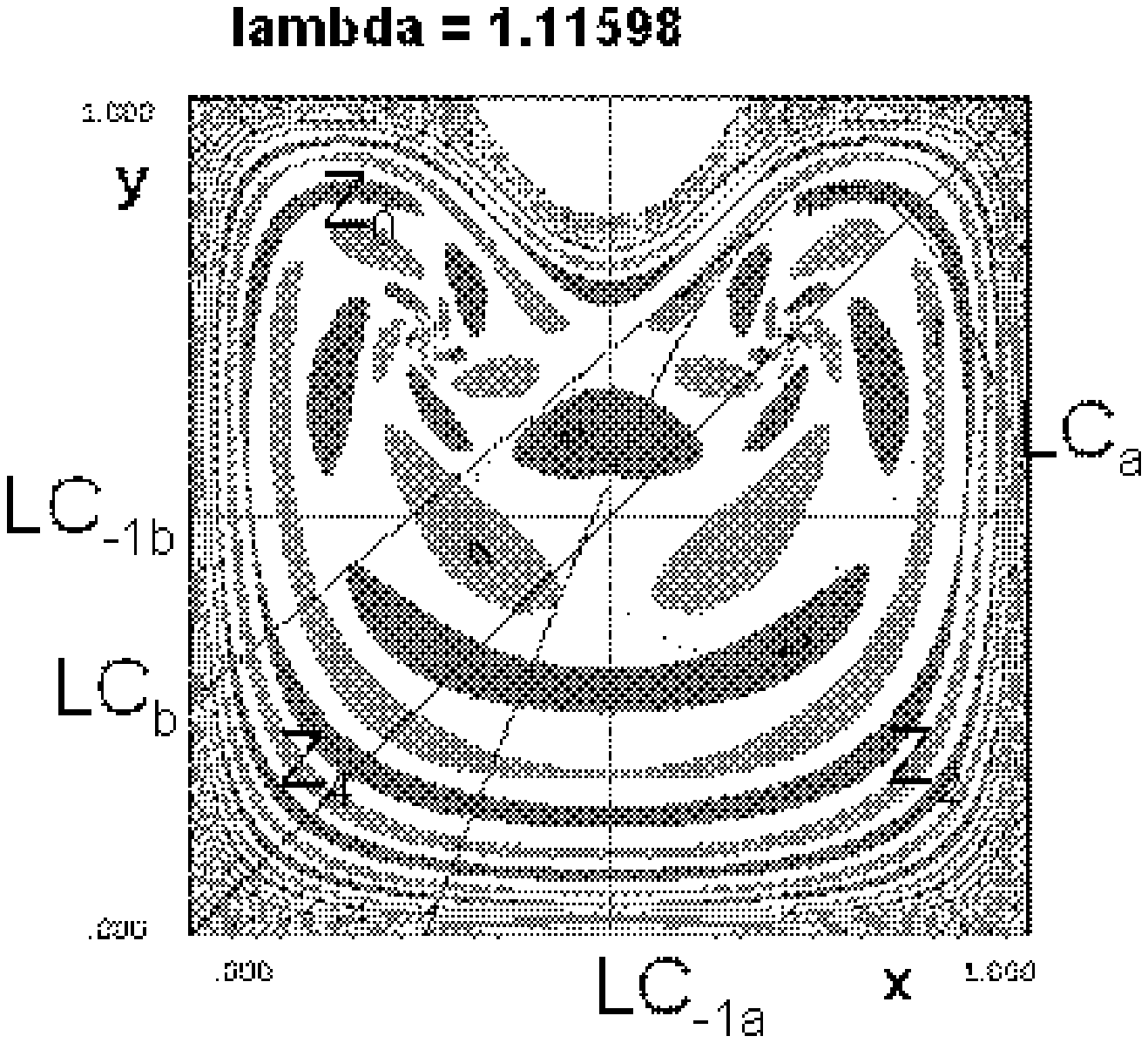}
      \caption[width=0.80]{Frequency locking on a period--29 cycle for the $ICC$ and Ne\"imark--Hopf bifurcation for the two period--3 attractive cycles. \label{fig21}}
   \end{minipage}
\end{figure}

\begin{figure}[t]	
   \begin{minipage}[c]{.50\linewidth}
      \includegraphics[width=0.80\textwidth]{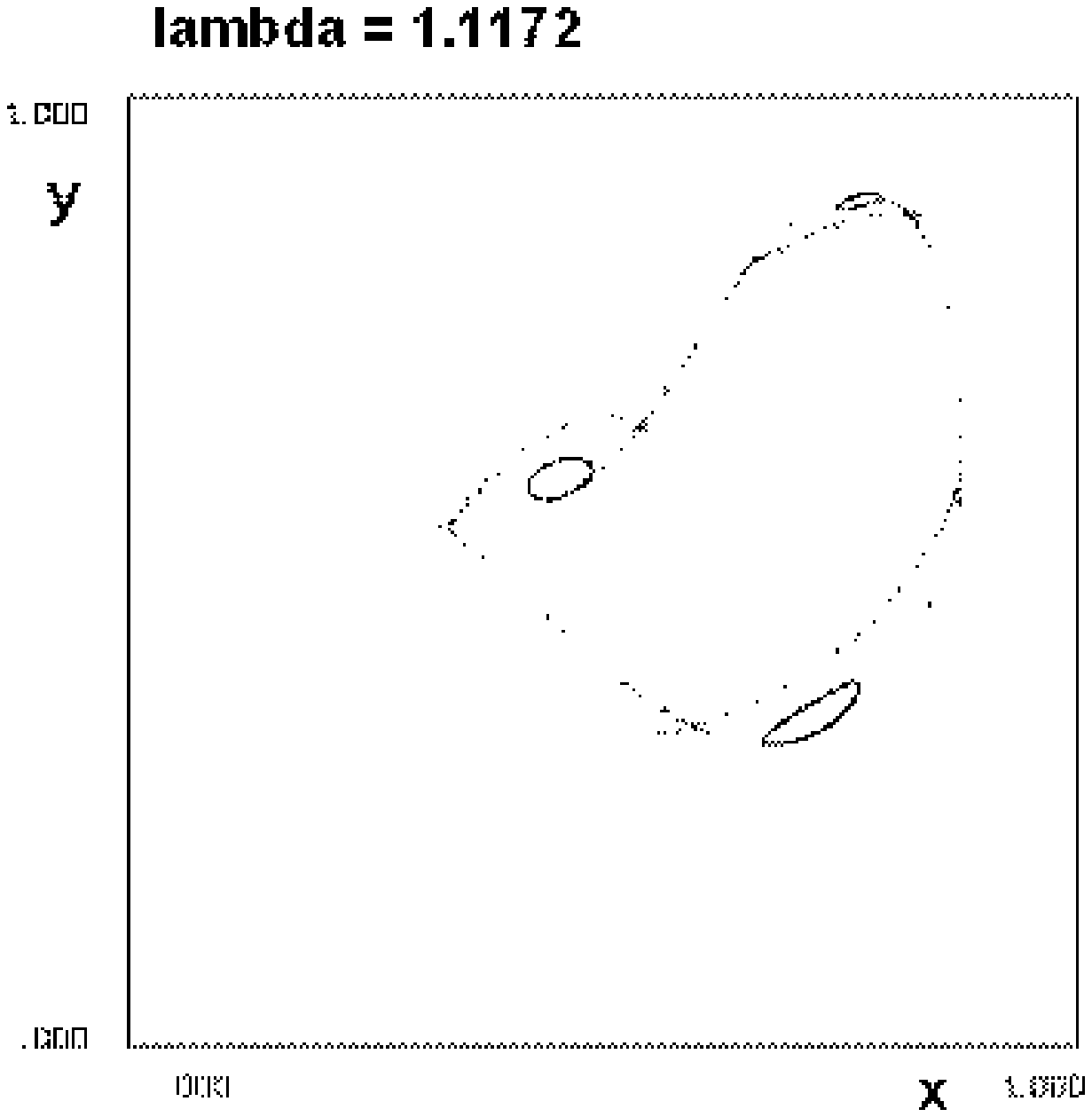}
      \caption[width=0.80]{Transient after that the $ICC$\\ $(C)$ has disappeared, the initial condition\\(.5,.51) gives rise to a trajectory, which\\follows the shape of the former $ICC$ $(C)$,\\ before reaching one of the period--3 $ICC$. \label{fig26}}
   \end{minipage} \hfill
   \begin{minipage}[c]{.50\linewidth}
      \includegraphics[width=0.80\textwidth]{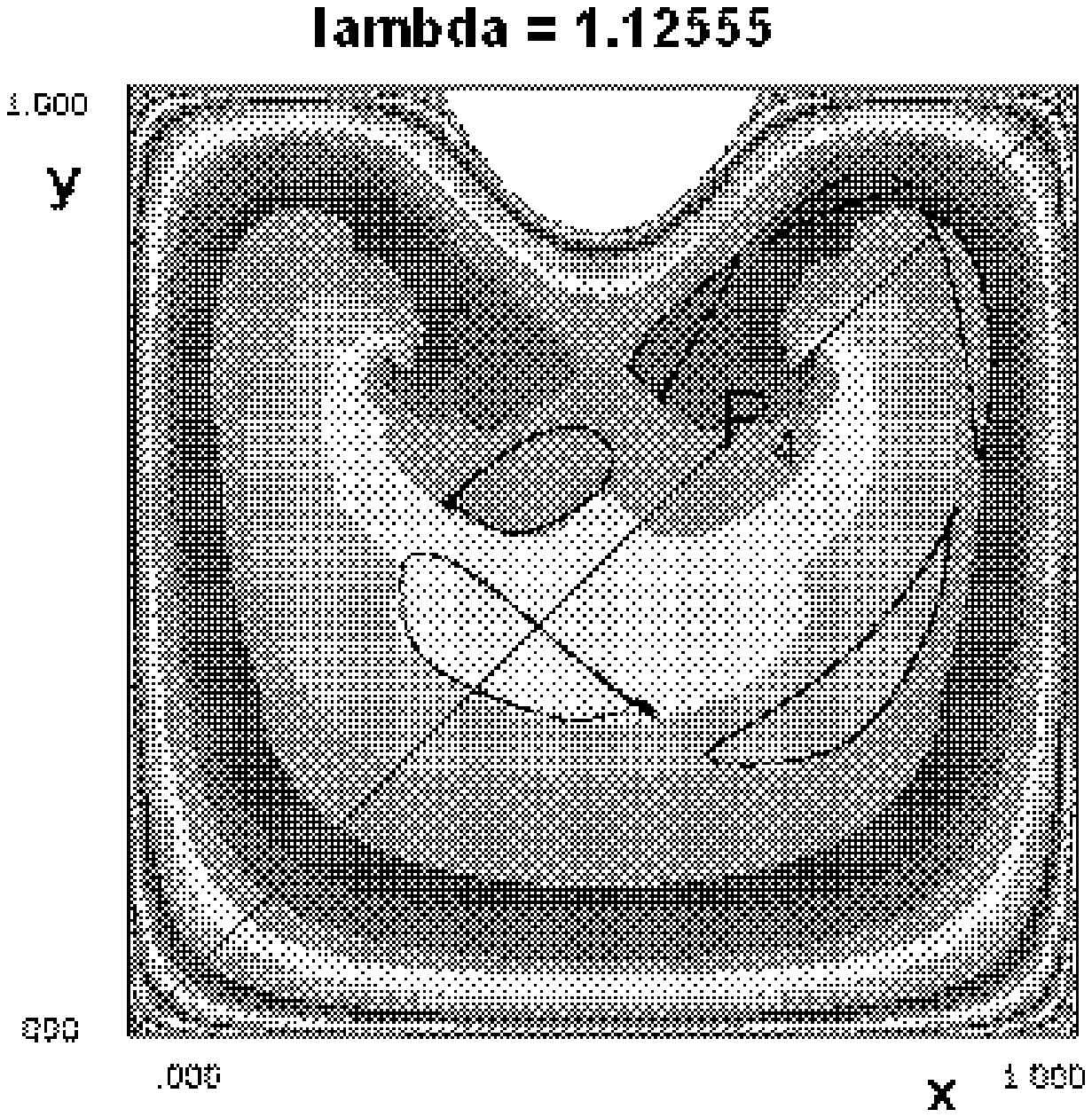}
      \caption[width=0.80]{Basins for the map T$^3$. It permits to see the basin of each curve, part of period--3 $ICC$. \label{fig27}}
   \end{minipage}
   \begin{center}
		\includegraphics[width=0.50\textwidth]{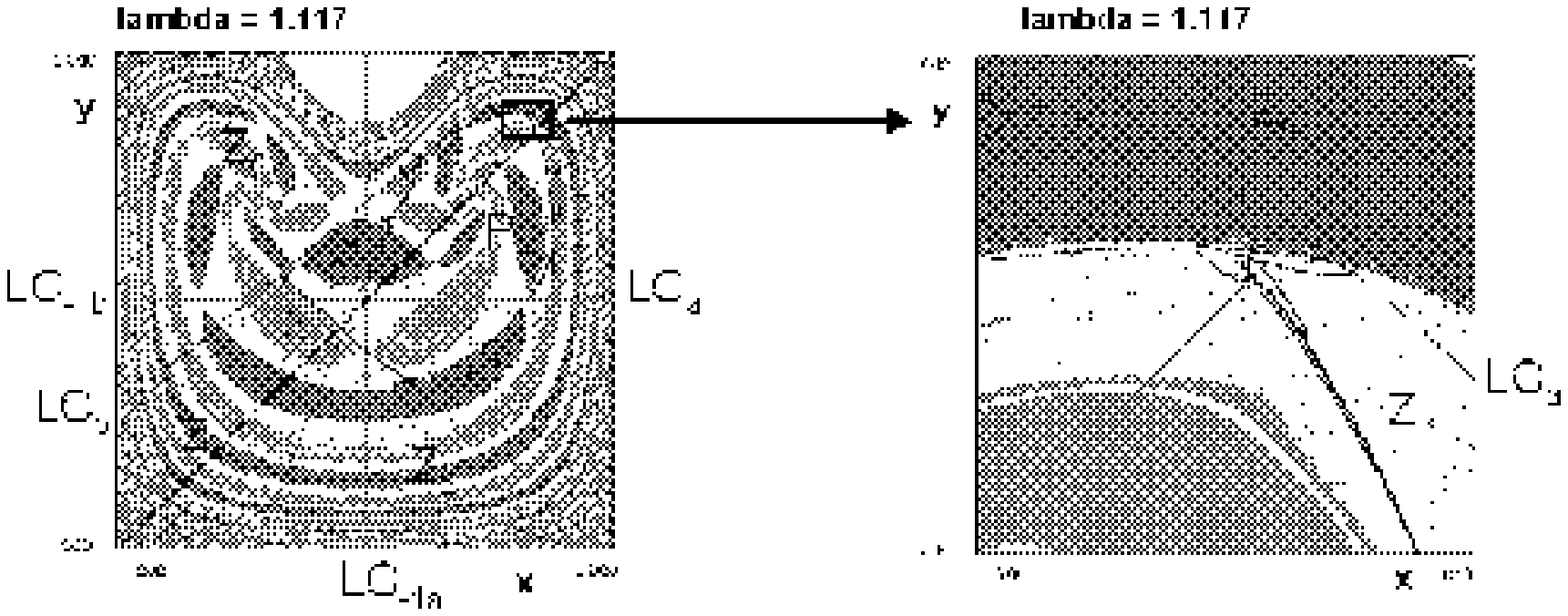}
	\end{center}
	\caption[width=0.80]{The curve $(C)$ becomes a $weakly\ chaotic\ ring$ when curles arise on it.}
	\label{fig23}
\end{figure}

\begin{figure}[t]
   \begin{minipage}[c]{.50\linewidth}
      \includegraphics[width=0.80\textwidth]{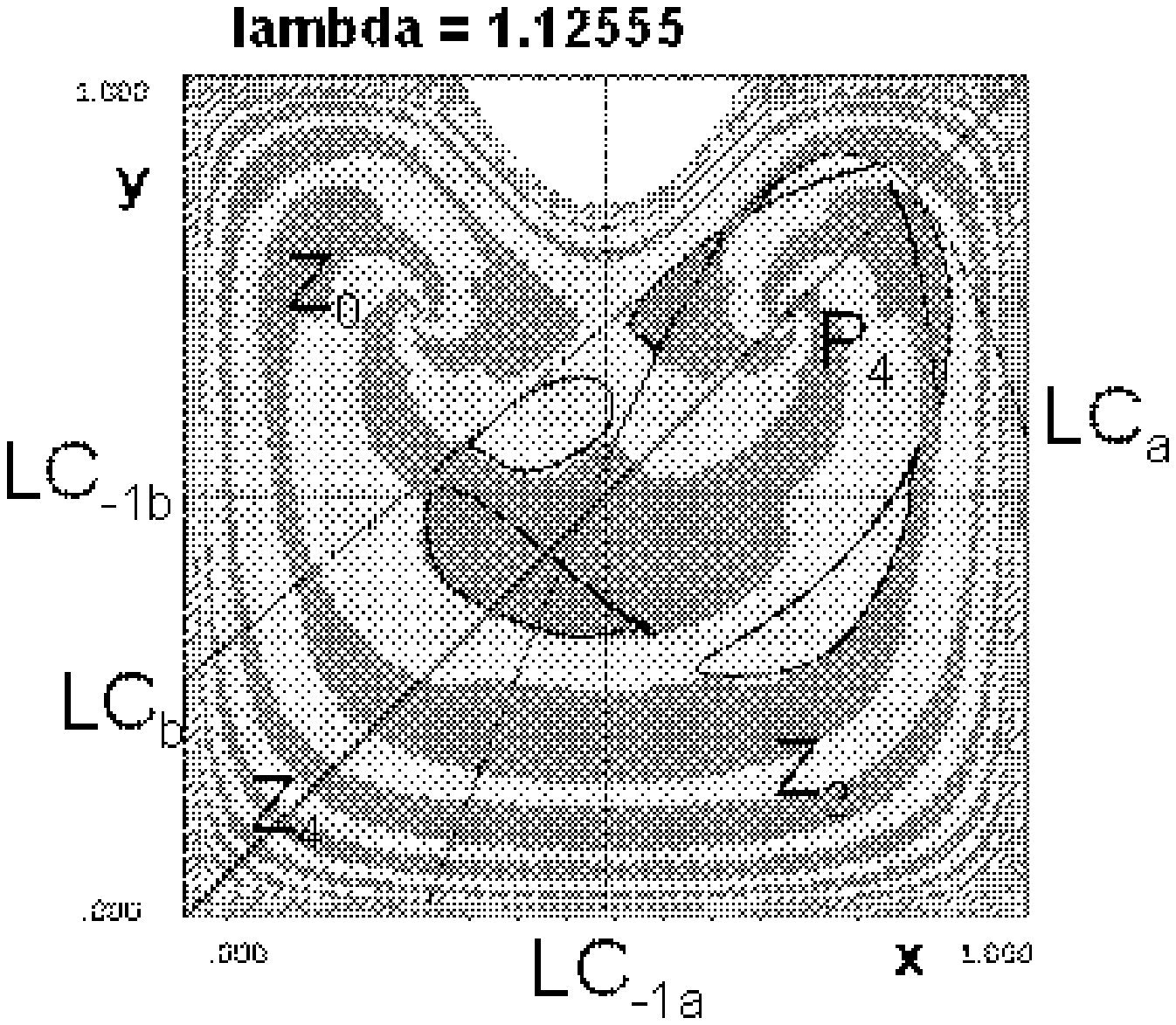}
      \caption[width=0.80]{Two 3--periodic stable $ICC$\\and their basin.
       \label{fig28}}
   \end{minipage} \hfill
   \begin{minipage}[c]{.50\linewidth}
      \includegraphics[width=0.80\textwidth]{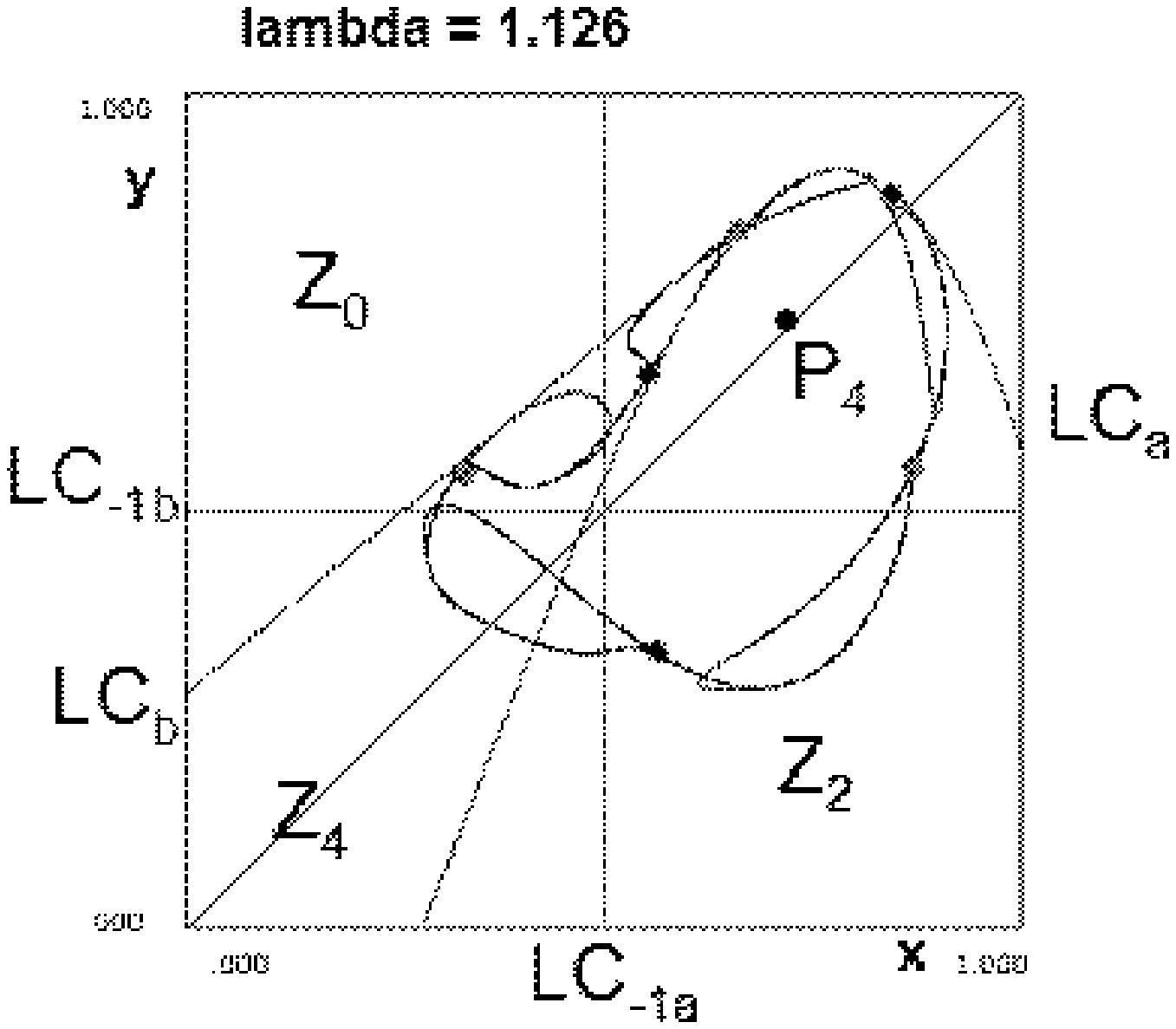}
      \caption[width=0.80]{After contact bifurcation, a single attractor exists. \label{fig29}}
   \end{minipage}
   \begin{minipage}[c]{.50\linewidth}
      \includegraphics[width=0.80\textwidth]{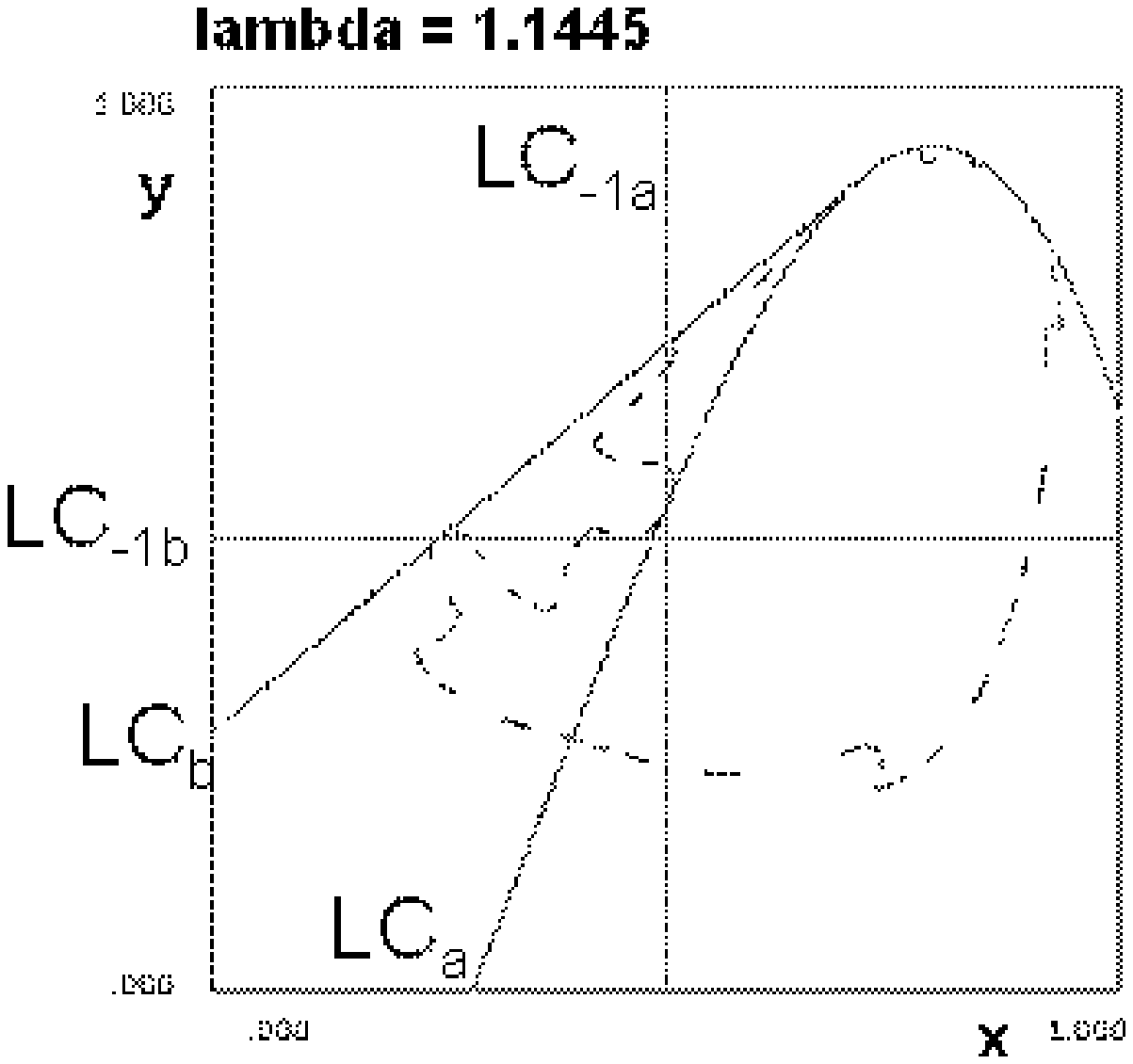}
      \caption[width=0.80]{Cyclic chaotic attractor.
       \label{fig30}}
   \end{minipage} \hfill
   \begin{minipage}[c]{.50\linewidth}
      \includegraphics[width=0.80\textwidth]{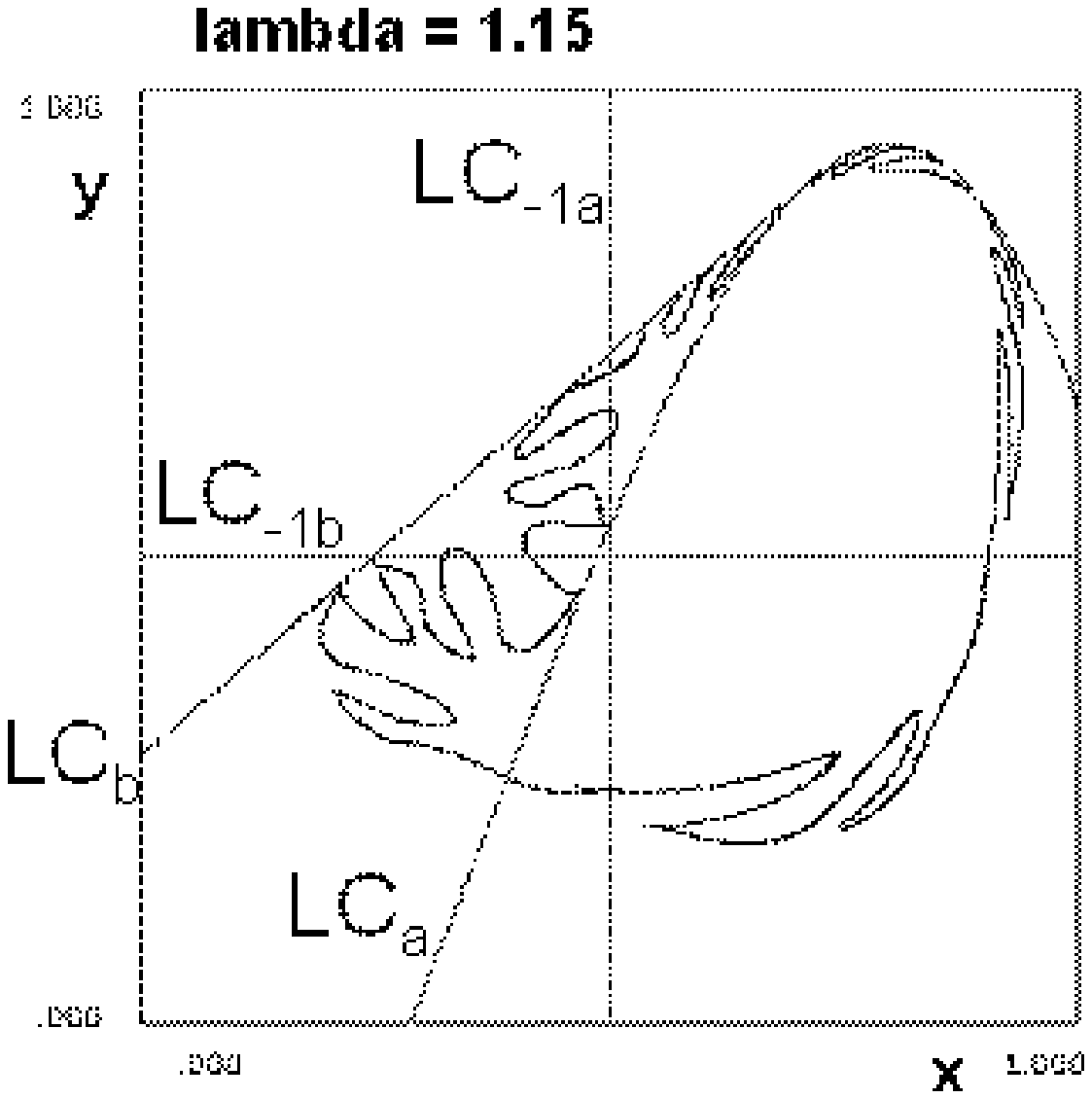}
      \caption[width=0.80]{Oscillations increase. \label{fig36}}
   \end{minipage}
\end{figure}

\begin{figure}[t]
   \begin{minipage}[c]{.50\linewidth}
      \includegraphics[width=0.80\textwidth]{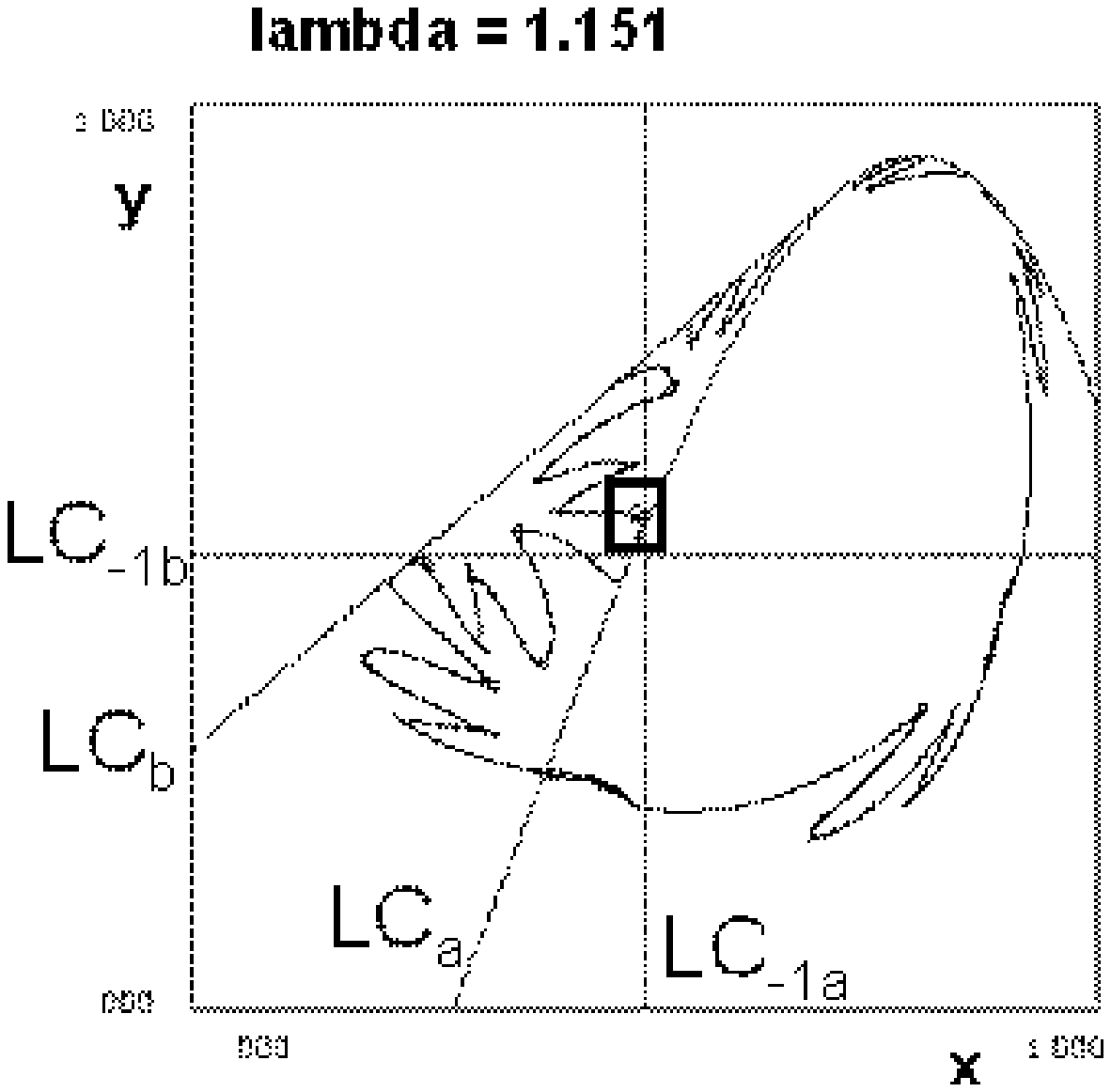}
      \caption[width=0.80]{Curles issued from heteroclinic tangencies.
       \label{fig40}}
   \end{minipage} \hfill
   \begin{minipage}[c]{.50\linewidth}
      \includegraphics[width=0.80\textwidth]{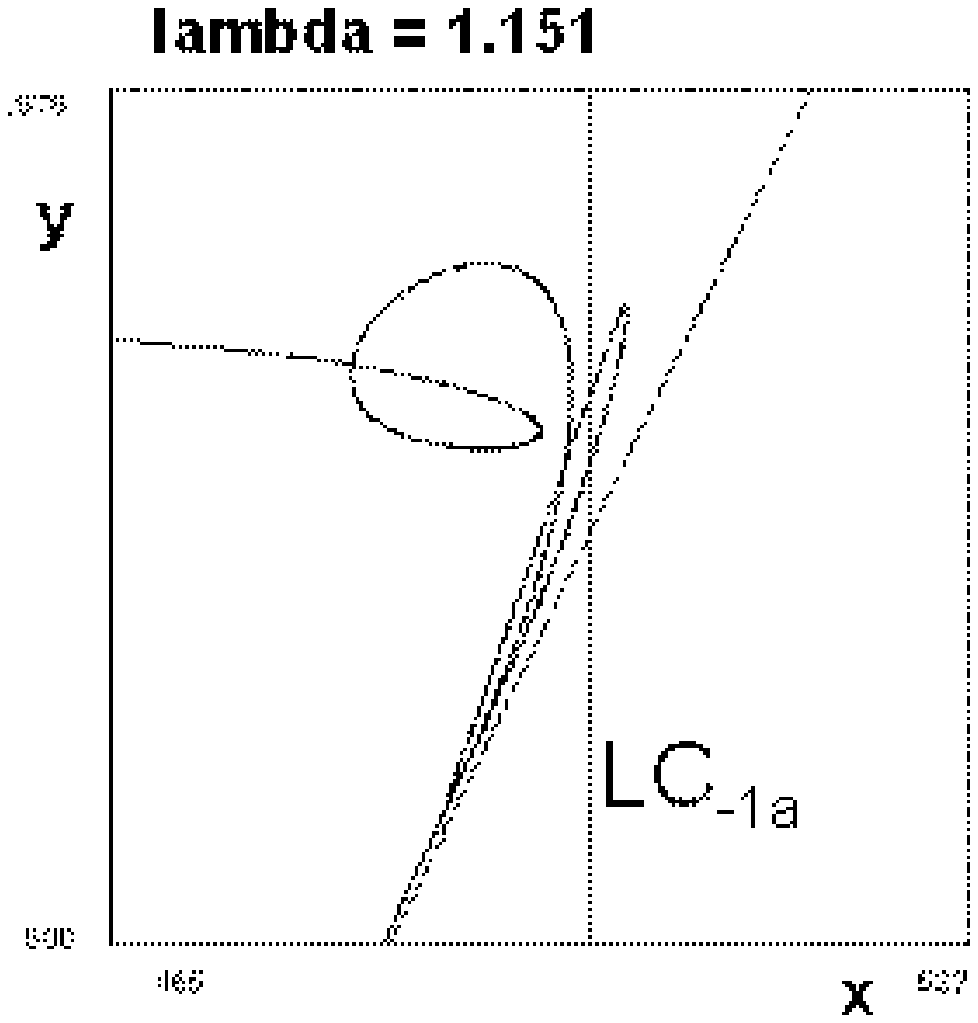}
      \caption[width=0.80]{Enlargment of Figure \ref{fig40} \label{fig37}}
   \end{minipage}
   \begin{minipage}[c]{.50\linewidth}
      \includegraphics[width=0.80\textwidth]{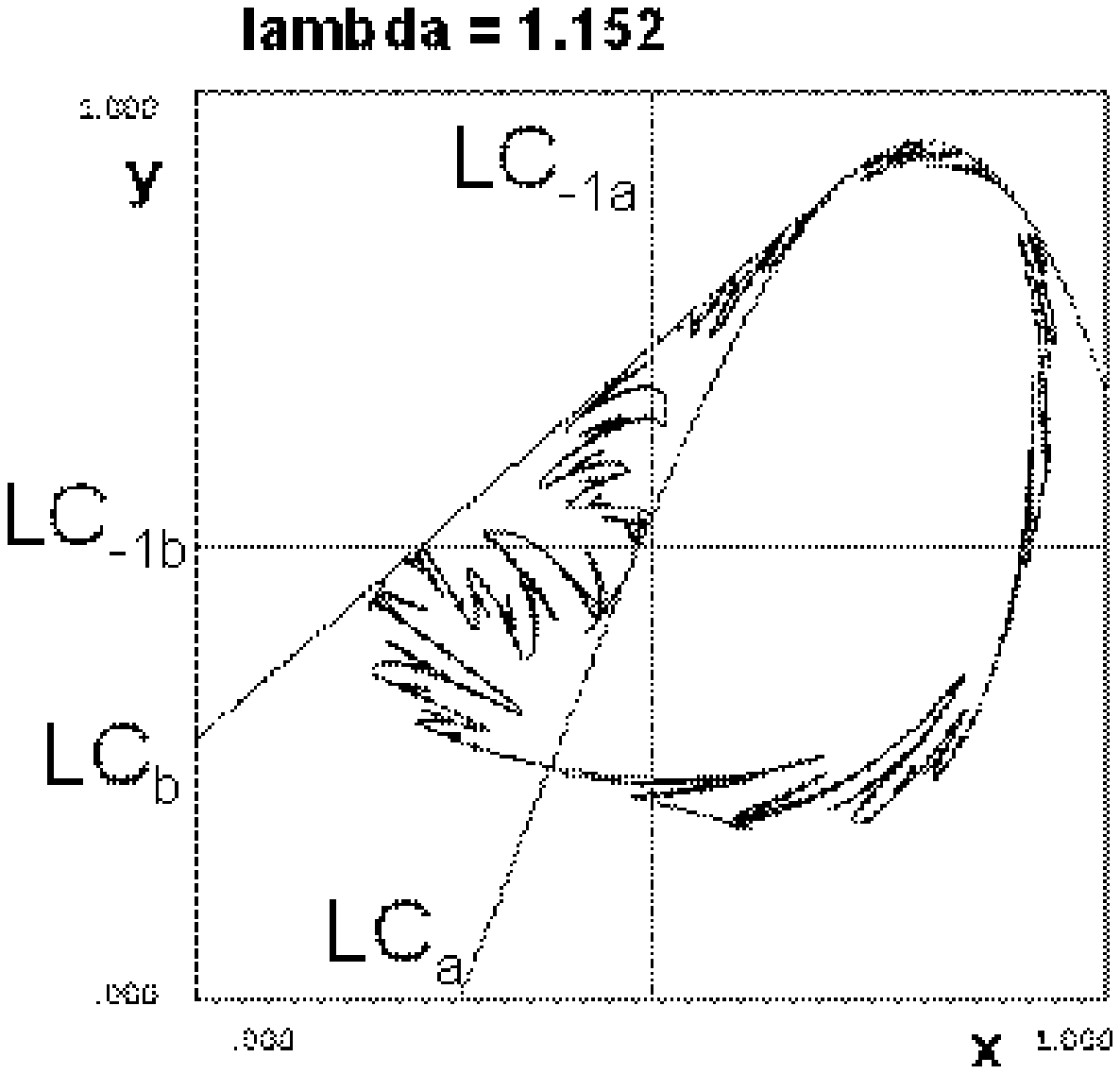}
      \caption[width=0.80]{Oscillations increase on the $WCR$. \label{fig38}}
   \end{minipage} \hfill
   \begin{minipage}[c]{.50\linewidth}
      \includegraphics[width=0.80\textwidth]{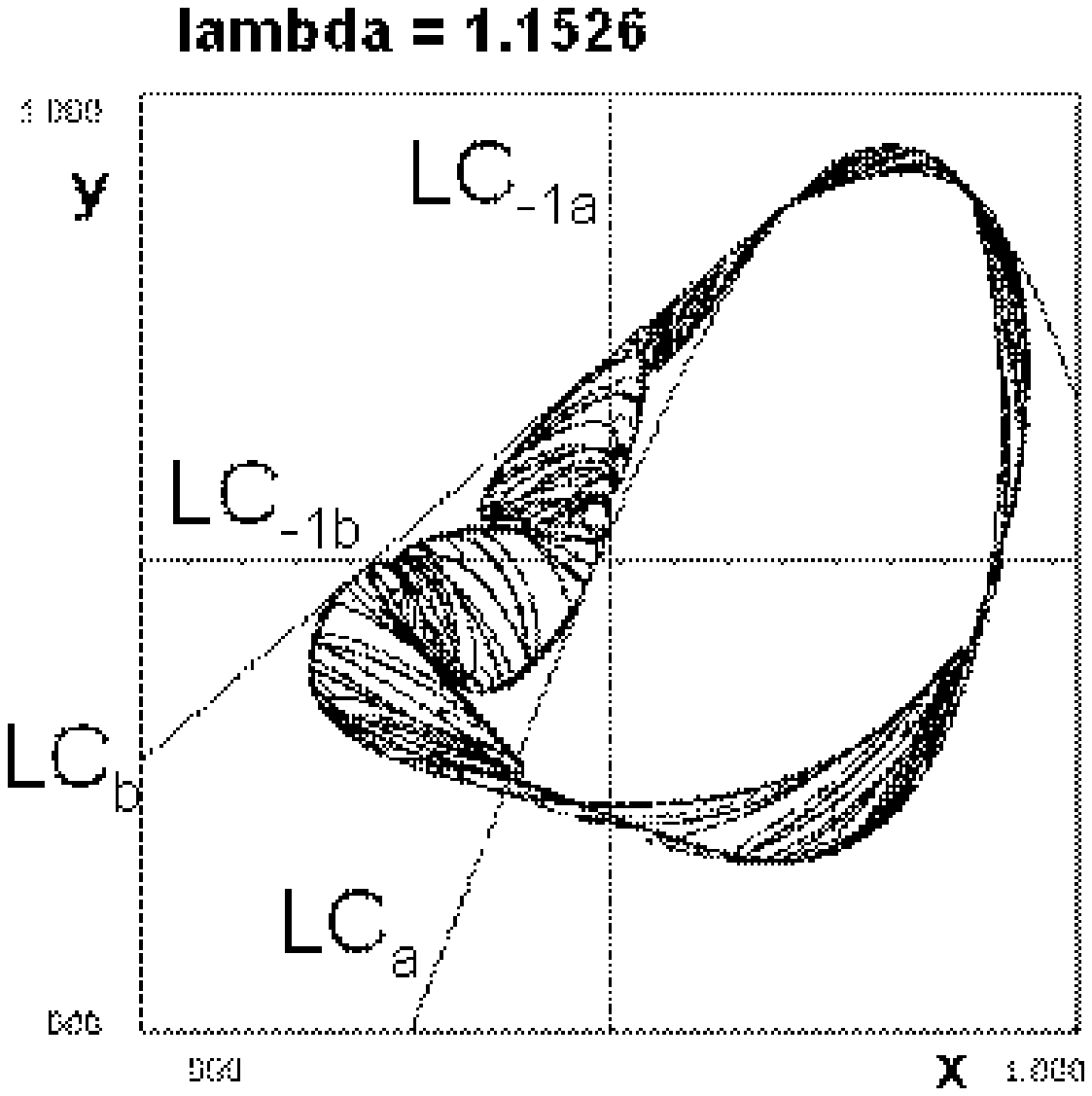}
      \caption[width=0.80]{Chaotic attractor. \label{fig43}}
   \end{minipage}
\end{figure}

\begin{figure}[t]
   \begin{minipage}[c]{.50\linewidth}
      \includegraphics[width=0.80\textwidth]{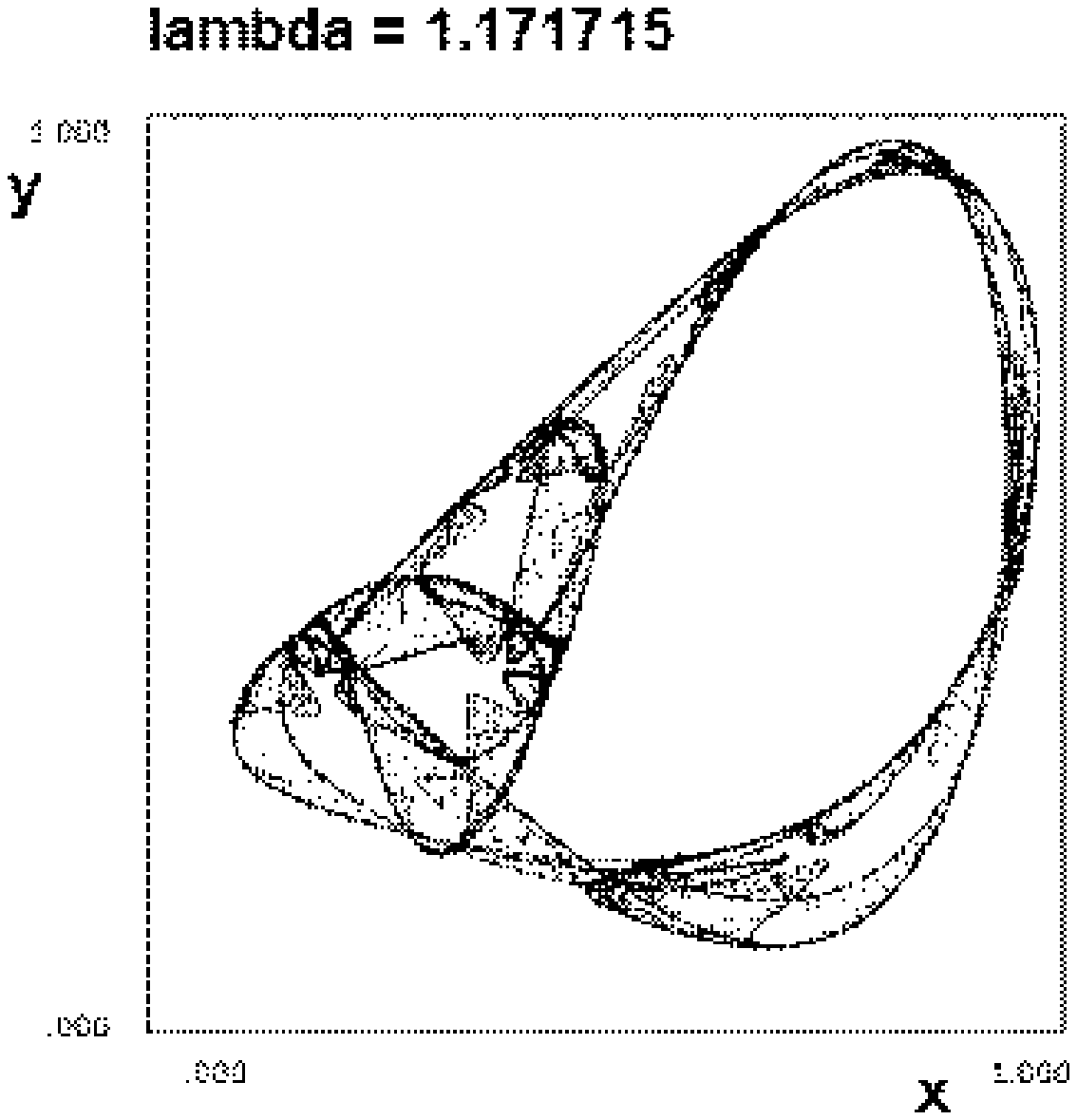}
      \caption[width=0.80]{Chaotic attractor. \label{fig44}}
   \end{minipage} \hfill
   \begin{minipage}[c]{.50\linewidth}
      \includegraphics[width=0.80\textwidth]{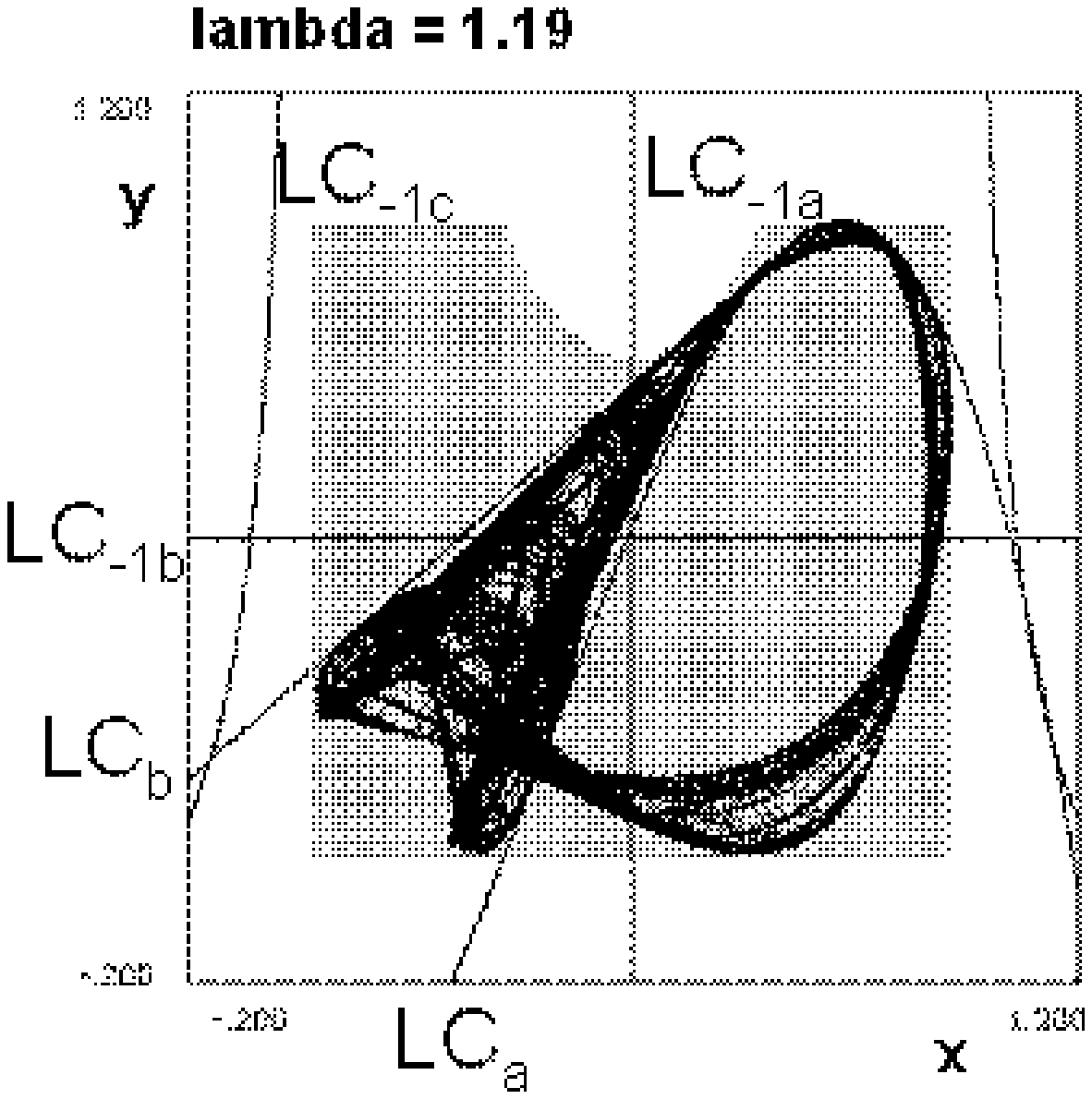}
      \caption[width=0.80]{Contact bifurcation of the chaotic attractor with its basin boundary. \label{fig46}}
   \end{minipage}
\end{figure}

We would like to thank Prof. Mira for helpful discussions, in particular, the Figure \ref{fig2a} is due to him. R. Lopez-Ruiz wishes to thank the Program Europa of CAI--CONSI+D (Zaragoza, Spain) for financial support.

\newpage

\end{document}